\documentclass[manuscript]{revtex4}
\usepackage{shuffle}
\usepackage{draft}
\usepackage{amsfonts}
\usepackage{mathrsfs}
\usepackage{amsmath}
\usepackage{amssymb}
\usepackage{bm}
\usepackage{feynmp-auto}
\usepackage[normalem]{ulem}
\usepackage{extarrows}
\usepackage{slashed}
\usepackage{float}
\usepackage{isodateo}
\usepackage{graphicx}
\usepackage{xcolor}
\usepackage[CJKbookmarks=true, bookmarksnumbered=true,bookmarksopen=true]{hyperref}
\hypersetup{colorlinks,%
	linkcolor=[rgb]{0,0.2,0.6}, %
	urlcolor=[rgb]{0,0.2,0.6}, %
	citecolor=[rgb]{0,0.2,0.6}}

\usepackage{indentfirst}

\graphicspath{{fig/}}
\usepackage{bm}
\usepackage{indentfirst}
\usepackage{tikz-cd}
\usetikzlibrary{patterns}
\usepackage{tikz}
\usetikzlibrary{arrows,shapes}
\usetikzlibrary{trees}
\usetikzlibrary{matrix,arrows} 				
\usetikzlibrary{positioning}				
\usetikzlibrary{calc,through}				
\usetikzlibrary{decorations.pathreplacing}  
\usepackage{pgffor}							

\usetikzlibrary{decorations.pathmorphing}	
\usetikzlibrary{decorations.markings}
\tikzset{
	photon/.style={decorate, decoration={snake}, draw=red},
	electron/.style={draw=blue, postaction={decorate},
		decoration={markings,mark=at position .55 with {\arrow[draw=blue]{>}}}},
	gluon/.style={decorate, draw=blue,
		decoration={coil,amplitude=4pt, segment length=4pt}} ,
	vector/.style={decorate, decoration={snake}, draw},
	provector/.style={decorate, decoration={snake,amplitude=2.5pt}, draw},
	antivector/.style={decorate, decoration={snake,amplitude=-2.5pt}, draw},
	fermion/.style={draw=black, postaction={decorate},
		decoration={markings,mark=at position .55 with {\arrow[draw=black]{>}}}},
	fermionbar/.style={draw=black, postaction={decorate},
		decoration={markings,mark=at position .55 with {\arrow[draw=black]{<}}}},
	fermionnoarrow/.style={draw=black},
	scalar/.style={dashed,draw=black, postaction={decorate},
		decoration={markings,mark=at position .55 with {\arrow[draw=black]{>}}}},
	scalarbar/.style={dashed,draw=black, postaction={decorate},
		decoration={markings,mark=at position .55 with {\arrow[draw=black]{<}}}},
	scalarnoarrow/.style={dashed,draw=black},
	electron/.style={draw=black, postaction={decorate},
		decoration={markings,mark=at position .55 with {\arrow[draw=black]{>}}}},
	bigvector/.style={decorate, decoration={snake,amplitude=4pt}, draw},
	background/.style={dashed,draw=black, postaction={decorate},
		decoration={markings,mark=at position 1 with {\arrow[draw=black]{<>}}}},
}

\tikzstyle{block} = [draw, rectangle, 
minimum height=3em, minimum width=6em]

\def\mb{\mathbf}
\def\ma{\mathcal}

\begin{document}

\title{A Type of Unifying Relation in (A)dS Spacetime}
	
\author{Yi-Xiao Tao}
\email{taoyx21@mails.tsinghua.edu.cn}
\affiliation{Department of Mathematical Sciences, Tsinghua University, Beijing 100084, China}
\author{Qi Chen}
\email{chenq20@mails.tsinghua.edu.cn}\thanks{The unusual ordering of authors instead of the standard alphabet ordering is for young researchers to get proper recognition of contributions under the current out-dated practice in China.}
\affiliation{Department of Physics, Tsinghua University, Beijing 100084, China}

\begin{abstract}
Unifying relations of amplitudes are elegant results in flat spacetime, but the research on these in (A)dS case is not very rich. In this paper, we discuss a type of unifying relation in (A)dS by using Berends-Giele currents. By taking the flat limit, we also get a semi-on-shell way to prove the unifying relations in the flat case. We also discuss the applications of our results in cosmology.	 
\end{abstract}

\maketitle

\tableofcontents	
\newpage	
\section{Introduction}
Scattering amplitudes are the most fundamental observables in quantum field theory (QFT), and a lot of on-shell methods at the tree level have been researched (see \cite{Britto:2004ap,Cheung:2015ota,Elvang:2013cua,Travaglini:2022uwo} and so on for a review). However, compared with the adequacy of the on-shell methods, other methods are less well studied. As a semi-on-shell method, Berends-Giele (BG) currents \cite{Berends:1987me} are popular because of their recursive character. In recent years, many techniques based on BG currents have been developed \cite{Mafra:2016ltu,Gomez:2022dzk,Armstrong:2022jsa,Chattopadhyay:2021udc,Du:2022vsw,Wu:2021exa,Frost:2020eoa,Mafra:2016mcc,Lee:2015upy,Mafra:2015gia,Mafra:2015vca,Mizera:2018jbh}. However, the research of BG currents in (A)dS is still very scarce. This paper explores BG currents' properties and potential applications in (A)dS. 

In flat spacetime, there are many elegant results for tree-level amplitudes. Unifying relations \cite{Cheung:2017ems} are such results. They reveal the implicit relations among different theories by some differential operators, which tells us that the theories that seem pretty different may have intrinsic relations. We hope these relations can be generalized to the tree-level correlation functions in (A)dS. It seems that some recursion relations or factorization methods \cite{Raju:2010by,Raju:2011mp,Raju:2012zr,Zhou:2018sfz,Baumann:2020dch,Arkani-Hamed:2015bza,Arkani-Hamed:2018kmz,Baumann:2019oyu,Baumann:2021fxj} are helpful in the proof of the unifying relations in (A)dS, which are the methods we use in the flat case. However, in dS spacetime, only the factorization methods for 4-pt correlation functions are understood very well \cite{Melville:2021lst,Goodhew:2021oqg}, thus it is still unknown whether they can be applied to this proof. Instead, BG currents recursion, as a semi-on-shell (semi-on-boundary) method, stands out because it is clearer than factorization methods in (A)dS \cite{Armstrong:2022jsa}. In this paper, we discuss the ``unifying relations" in (A)dS between gluons and conformally coupling scalars (minimal coupled with gluons), which can be thought as a special case of Yang-Mills scalar (YMS) theory by setting the second color group to be the $U(1)$ group (the first color group is the gauge group), by using the recursion properties of BG currents. We find that this type of unifying relations also holds in (A)dS and develop some methods based on BG currents. By the way, we can also prove the ``unifying relations" for flat spacetime by taking the flat limit. 

Based on our conclusion, we want to discover some potential cosmology applications. It is widely held that at the beginning of our universe, it underwent an exponential expansion called cosmic inflation, and its background spacetime can be viewed as the dS space. During the period of inflation, the quantum fluctuations for all possible fields provide a source to generate the non-gaussianity and seed the Large Scale Structure (LSS) at present. In other words, the fluctuations of particles imprint some signals in the sky today, which allows us to research the history of the primordial universe. Such fascinating information can be abstracted from the cosmological correlation functions\cite{Maldacena:2002vr,Arkani-Hamed:2015bza}. So far, the 4-pt correlation functions are the most valuable ones in cosmic experiments, and it is complicated to calculate \cite{Arkani-Hamed:2018kmz,Sleight:2019hfp,Baumann:2019oyu,Baumann:2020ksv,Baumann:2020dch,Baumann:2022jpr}. In this paper, we present a new aspect to understand the relations among different tree-level cosmological correlation functions by unifying relations, and give some examples at 4-pt level.

The paper is organized as follows. In section \ref{sec:unifyingflat}, we briefly review the remarkable unifying relations. In section \ref{sec:BGcurrent}, we review the concepts of BG currents and introduce the methods to calculate them. In section \ref{sec:unifyingads}, we give some examples of a type of unifying relations for BG currents in (A)dS and then discuss the general form. In section \ref{sec:applications}, we talk about some applications of our results in cosmology.

\section{Unifying Relations in Flat Spacetime}
\label{sec:unifyingflat}
The unifying relations are first discovered in flat spacetime \cite{Cheung:2017ems} and are found to be consistent with Cachazo-He-Yuan (CHY) formalism  \cite{Zhou:2018wvn}. Let us review the unifying relations in \cite{Cheung:2017ems} very briefly.

There are many theories in QFT, such as Yang-Mills (YM) theory, the bi-adjoint scalar (BS) theory, the special Galileon (SG) theory, and the Yang-Mills scalar (YMS) theory. Moreover, they seem to be quite different from each other formally. However, their amplitudes at the tree level can be related by using some differential operators. These operators need to preserve on-shell kinematics and gauge invariance of the amplitudes. Here we write down the operators directly:
\ie
&\mathcal{T}_{ij}=\partial_{\epsilon_{i}\epsilon_{j}}\\
&\mathcal{T}_{ijk}=\partial_{k_{i}\epsilon_{j}}-\partial_{k_{k}\epsilon_{j}}\\
&\mathcal{T}[\alpha]=\mathcal{T}_{\alpha_{1}\alpha_{n}}\cdot\prod_{i=2}^{n-1}\mathcal{T}_{\alpha_{i-1}\alpha_{i}\alpha_{n}}\\
&\mathcal{L}_{i}=\sum_{j}k_{i}k_{j}\partial_{k_{j}\epsilon_{i}}\\
&\mathcal{L}=\prod_{i}\mathcal{L}_{i}.
\fe
Here $k_{i}$ is the momentum of the $i$-th particle, and $\epsilon_{i}$ is the polarization vector of the $i$-th particle. Using these operators, we can transform the tree-level amplitudes of some theories into other theories. These transformations are concluded as ``unifying relations". We can write down these unifying relations explicitly for color-ordered amplitudes:
\ie\label{uni}
A_{\rm YMS\,}&=\mathcal{T}[i_{1}j_{1}]\mathcal{T}[i_{2}j_{2}]\cdots\mathcal{T}[i_{n}j_{n}]\cdot A_{\rm YM\,}\\
A_{\rm BS\,}&=\mathcal{T}[i_{1}i_{2}\cdots i_{n}]\cdot A_{\rm YM\,}\\
A_{\rm NLSM\,}&=\mathcal{L}\cdot\mathcal{T}[i_{1}i_{n}]\cdot A_{\rm YM\,}.
\fe
By replacing $A_{\rm YM\,}$ with BI amplitudes $A_{\rm BI\,}$ or extended gravity amplitudes $A_{\rm G\,}$, we will get other unifying relations:
\ie\label{uni2}
A_{\rm DBI\,}&=\mathcal{T}[i_{1}j_{1}]\mathcal{T}[i_{2}j_{2}]\cdots\mathcal{T}[i_{n}j_{n}]\cdot A_{\rm BI\,}\\
A_{\rm NLSM\,}&=\mathcal{T}[i_{1}i_{2}\cdots i_{n}]\cdot A_{\rm BI\,}\\
A_{\rm SG\,}&=\mathcal{L}\cdot\mathcal{T}[i_{1}i_{n}]\cdot A_{\rm BI\,}\\
A_{\rm EM\,}&=\mathcal{T}[i_{1}j_{1}]\mathcal{T}[i_{2}j_{2}]\cdots\mathcal{T}[i_{n}j_{n}]\cdot A_{\rm G\,}\\
A_{\rm YM\,}&=\mathcal{T}[i_{1}i_{2}\cdots i_{n}]\cdot A_{\rm G\,}\\
A_{\rm BI\,}&=\mathcal{L}\cdot\mathcal{T}[i_{1}i_{n}]\cdot A_{\rm G\,}.
\fe
Here the ``EM" is the Einstein-Maxwell theory. We need to point out that the particles denoted by indices in each $\mathcal{T}[\alpha]$ are in the same trace after $\mathcal{T}[\alpha]$ acting on a certain amplitude. By the way, $\ma{T}[ij]=\mathcal{T}_{ij}$ is called ``trace operator". For example,
\ie
A_{\rm YMS\,}(\phi_{1}\phi_{2},\phi_{3}\phi_{4},\cdots,\phi_{2n-1}\phi_{2n})=\mathcal{T}[12]\mathcal{T}[34]\cdots\mathcal{T}[2n-1,2n]\cdot A_{\rm YM\,}(g_{1},g_{2},\cdots,g_{2n-1},g_{2n}).
\fe
Another important thing is that we can also use this relation to get a mixed amplitude. For example,
\ie\label{mix}
A_{\rm YMS\,}(\phi_{1}\phi_{2},g_{3},g_{4})=\mathcal{T}[12]\cdot A_{\rm YM\,}(g_{1},g_{2},g_{3},g_{4}).
\fe
The proof of these relations is based on the on-shell factorization. 

An important corollary of the unifying relations is the relations between the YM theory and the scalar with a minimal coupling with gluons. In this case, there are no $\phi^{3}$ and $\phi^{4}$ vertices, and scalars have only one color index. This case is equivalent to the case that we choose the second color group of the YM scalar to be the $U(1)$ group (the first color group is the gauge group). We define that the scalars separated from other scalars by gluon propagators are in the same trace, which is just a generalization of the trace for scalars in the general YMS theory. Obviously, each trace can only have 2 scalars. Scalars have only one color index means that we cannot distinguish different trace structures of a certain set of scalars by Feynman rules, and we have to sum over all possible trace structures to get the color-ordered amplitudes we want. Hence the unifying relations for the YM theory and the YMS theory can be modified as follows so that they hold for the scalar theory with a minimal coupling with gluons:
\ie\label{cor}
\ma{T}^{X}A_{YM}(g_{1},g_{2}\cdots g_{n})=A_{S}(\phi_{X},g_{\{1,2,\cdots,n\}\backslash X})
\fe
 where $A_{S}$ is the amplitude of the scalar theory with a minimal coupling with gluons, and $\phi_{X}$ means that the letters in $X\subseteq \{1,2,3,\cdots,n\}$ correspond to scalar legs in $A_{S}$ and so as $g_{\{1,2,\cdots,n\}\backslash X}$ for gluon legs. Here $\ma{T}^{X}$ means we pair the letters in the word $X$ to get a product of $\ma{T}[ij]$ based on this pairing method and then sum over all pairing methods. For example, we have $\ma{T}^{1234}=\ma{T}[12]\ma{T}[34]+\ma{T}[13]\ma{T}[24]+\ma{T}[14]\ma{T}[23]$. Note that $|X|$, the length of the word $X$, must be even, or we will get zero because in such a theory we cannot have an amplitude with odd number scalar legs. In the following discussion, we will focus on the proof of this corollary in (A)dS by using BG currents, and we also use ``unifying relation" to denote the corollary \eqref{cor} in addition to \eqref{uni} and \eqref{uni2} in the following discussion.

In this paper, we will not discuss the operator $\mathcal{L}$, so we skip the discussion about that.  For a more comprehensive introduction, see \cite{Cheung:2017ems}.

\section{BG Currents and Recursion}
\label{sec:BGcurrent}
In this section, we will talk about the BG currents and recursions both in the flat and (A)dS cases. They can both be derived from the classical equation of motion \cite{Mizera:2018jbh,Frost:2020eoa} and are connected by taking the flat limit, which will be introduced in this section.

\subsection{BG Currents in Flat Spacetime}
\label{subsec:BGflat}
In this subsection, we will review some properties of BG currents in flat spacetime. We will mainly follow the methods in \cite{Mizera:2018jbh} but with different normalizations.

The definition of $n$-pt BG currents is $n$-pt tree amplitudes with an external leg being off-shell. It is well-known that BG currents can be computed from the classical equation of motion. Here we give the calculation of the BG currents of YM theory. From now on, we will take the Lorenz gauge. First of all, we write down the Lagrangian of YM theory (here we use the matrix formalism):
\ie
\mathcal{L}_{\rm YM\,}=\frac{1}{4}\Tr(\mathbb{F}_{\mu\nu}\mathbb{F}^{\mu\nu}),
\fe
where $\mathbb{F}_{\mu\nu}=\partial_{\mu}\mathbb{A}_{\nu}-\partial_{\nu}\mathbb{A}_{\mu}-i[\mathbb{A}_{\mu},\mathbb{A}_{\nu}]$ and $\nabla_{\mu}=\partial_{\mu}-i[\mathbb{A}_{\mu},\cdot]$. The equation of motion is
\ie\label{glueom}
\square\mathbb{A}^{\nu}(x)=i[\mathbb{A}_{\mu}(x),\partial^{\mu}\mathbb{A}^{\nu}(x)]+i[\mathbb{A}_{\mu}(x),\mathbb{F}^{\mu\nu}(x)].
\fe
We introduce the perturbiner expansion ansatz first:
\ie\label{an}
\mathbb{A}^{\mu}&:=\sum_{P}\mathcal{A}_{P}^{\mu}T^{P}e^{ik_{P}x}=\sum_{i}\mathcal{A}_{i}^{\mu}T^{a_{i}}e^{ik_{i}x}+\sum_{i,j}\mathcal{A}_{ij}^{\mu}T^{a_{i}}T^{a_{j}}e^{ik_{ij}x}+\cdots\\
\mathbb{F}^{\mu\nu}&:=i\sum_{P}\mathcal{F}^{\mu\nu}_{P}T^{P}e^{ik_{P}x}.
\fe
Here $P$ goes over all non-empty words. For $\mathbb{A}_{\mu}$ is in adjoint representation, we have the following shuffle identities \cite{Kleiss:1988ne,Lee:2015upy,Berends:1988zn}:
\ie\label{shuffle}
\ma{A}_{P\shuffle Q}^{\mu}=\ma{F}_{P\shuffle Q}^{\mu\nu}=0,\ \ P,Q\neq\varnothing.
\fe
where $\shuffle$ means we sum over all the permutations of the labels in $P\cup Q$ with the ordering of labels in $P$ and $Q$ preserved respectively. Then one can understand the perturbiner expansion ansatz \eqref{an} as ``power series" of the structure constant $f_{abc}$ of the corresponding Lie algebra. 

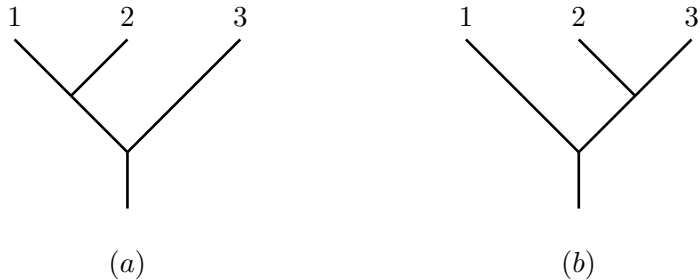
\begin{figure}[H]
\centering
\begin{tikzpicture}[line width=1pt,scale=1.5]
\draw[fermionnoarrow](2,-1)--(3,0);
\draw[fermionnoarrow](1.5,-0.5)--(2,0);
\draw[fermionnoarrow](1.5,-0.5)--(1,0);
\draw[fermionnoarrow](2,-1)--(1.5,-0.5);
\draw[fermionnoarrow](2,-1)--(2,-1.5);
\node at (1,0.2) {1};
\node at (2,0.2) {2};
\node at (3,0.2) {3};
\node at (2,-2) {$(a)$};
\begin{scope}[shift={(4,0)}]
\draw[fermionnoarrow](2,-1)--(3,0);
\draw[fermionnoarrow](2.5,-0.5)--(2,0);
\draw[fermionnoarrow](1.5,-0.5)--(1,0);
\draw[fermionnoarrow](2,-1)--(1.5,-0.5);
\draw[fermionnoarrow](2,-1)--(2,-1.5);
\node at (1,0.2) {1};
\node at (2,0.2) {2};
\node at (3,0.2) {3};
\node at (2,-2) {$(b)$};
\end{scope}
\end{tikzpicture}
\caption{Binary representations for Lie monomials. $(a)$ is the diagrammatic representation for [[1,2],3] and $(b)$ represents the Lie algebra structure for [1,[2,3]].}
\label{fig:binarytree}
\end{figure}

These expansion is called ``color-stripped" for the coefficients $\mathcal{A}_{P}^{\mu}$ and $\mathcal{F}^{\mu\nu}_{P}$ do not have any color degrees of freedom. We can construct color-ordered amplitudes by these coefficients. After substituting the ansatz and shuffle identities to the equation of motion, we have
\ie\label{tree}
-s_{P}\mathcal{A}_{P}^{\mu}&=\sum_{P=XY}[\mathcal{A}_{X}^{\mu}(k_{X}\cdot \mathcal{A}_{Y})+\mathcal{A}_{X\nu}\mathcal{F}_{Y}^{\mu\nu}-(X\leftrightarrow Y)]\\
\mathcal{F}_{Y}^{\mu\nu}&=k_{Y}^{\mu}\mathcal{A}_{Y}^{\nu}-k_{Y}^{\nu}\mathcal{A}_{Y}^{\mu}-\sum_{Y=RS}(\mathcal{A}_{R}^{\mu}\mathcal{A}_{S}^{\nu}-\mathcal{A}_{S}^{\mu}\mathcal{A}_{R}^{\nu}).
\fe
where $1/s_{P}$ is the propagator of the off-shell leg and the sum over $P=JK$ represents the sum over divisions which include all the order-preserving ways of splitting the $P$ into $JK$, like $P=12345$ and $J=123$, $K=45$. Note that the sum in \eqref{tree} can be represented as the sum of all possible binary trees consistent with $P$, where different binary trees correspond to different deconcatenation methods; see \cite{Mafra:2020qst,Frost:2020eoa}. Here `` consistent with $P$" means that the letters appear in the Lie monomial corresponds to a certain binary tree is the same as the letters appear in $P$ and the order of letters (from left to right) in the Lie monomial is the same as the order of letters in $P$ (The phrase ``Lie monomial" will be explained later). In \cite{Frost:2020eoa}, they use a different notation. One of the advantages of their notation is that we can divide our currents into many binary trees, and each tree relates to a Lie monomial \cite{garsia1990combinatorics}, which brings us many conveniences. This correspondence can be obtained by replacing the generators $T^{a_{i}}$ by the indices $i$ and then substituting our ansatz \eqref{an} and shuffle identities \eqref{shuffle} into \eqref{glueom}, after which we will get some Lie monomials like [[1,2],3]. This Lie monomial corresponds to such a binary tree: 1 and 2 are two different branches originating from the vertex; then 1 and 2, as a whole, together with 3 are also two different branches from a new vertex. These can be diagrammatically represented by the binary tree FIG. \ref{fig:binarytree} (a), and the deconcatenation $P=XY,\ X=RS$ for this binary tree is $P=123,\ X=12,\ Y=3,\ R=1,\ S=2$. For a more detailed review, see \cite{Frost:2020eoa}.

We need to point out that in \eqref{tree} the terms summed over once involve the information of 3-pt vertices, while the terms summed over twice, for example, $\ma{A}^{\mu}_{R}(\ma{A}_{X}\cdot\ma{A}_{S})$, involves 4-pt vertices since any two adjacent vertices of a binary tree has 4 lines connecting to other vertices so that it can be thought as a 4-pt vertex effectively. We impose the one-particle states as the initial conditions: 
\ie
\mathcal{A}_{i}^{\mu}=\epsilon_{i}^{\mu},
\fe
where the polarization vectors $\epsilon_{i}^{\mu}$ satisfy the transversality condition $\epsilon_{i}\cdot k_{i}=0$.

We can construct amplitudes from BG currents, which can be realized by moving the off-shell leg to be on-shell. For YM color-ordered amplitudes $A_{\rm YM\,}$, we have
\ie
A_{\rm YM\,}(Pn)=s_{P}\mathcal{A}_{P}\cdot \mathcal{A}_{n}\ \ (s_{P}\rightarrow0).
\fe

For BS theory, we can also obtain its BG currents:
\ie
\Phi_{P|Q}=\frac{1}{s_{P}}\sum_{P=RS}\sum_{Q=TU}[\Phi_{R|T}\Phi_{S|U}-(R\leftrightarrow S)],
\fe
and the initial conditions $\Phi_{i|j}=\delta_{ij}$. One can also see \cite{Mafra:2016ltu,Frost:2020eoa} for a more compact formalism. The double color-ordered amplitude of BS theory can also be constructed by BG currents:
\ie
m(Pn|Qn)=s_{P}\Phi_{P|Q}\Phi_{n|n}\ \ (s_{P}\rightarrow0).
\fe

We should point out an interesting thing. By counting the number of $\epsilon_{i}\cdot k_{j}$ and $\epsilon_{i}\cdot\epsilon_{j}$, we find that
\ie\label{start}
\Phi_{P|Q}=\mathcal{T}[Qn]\cdot (\mathcal{A}_{P}\cdot \epsilon_{n})
\fe
is correct even for an arbitrary binary tree up to an overall constant. Here we require that the length of the word $P$ is the same as $Qn$. Note that here the $n$-th particle is still off-shell, and the vector $\epsilon_{n}$ is only a formal polarization vector which is used for the convenience of allowing us to use the trace operator $\ma{T}[Qn]$. Then \eqref{start} is just a relation for currents, and it is equivalent to the relation for corresponding amplitudes since the unifying relations for amplitudes can easily be derived from the unifying relations for currents and vice versa. This fact implies that we can use BG currents to prove some of the unifying relations for amplitudes by proving the relations for currents of every binary tree, which means we find a semi-on-shell way to prove the unifying relations for amplitudes. This is the starting point of our work.

\subsection{BG Currents in (A)dS}\label{BGAdS}
Now we tend to focus on the BG currents in curved spacetime. Though BG currents themselves were proposed in flat spacetime \cite{Berends:1988zn} for the first time, we can also generalize it to (A)dS \cite{Armstrong:2022jsa}. We can define BG currents as $n$-pt tree-level correlation functions with all but one external leg on the spacetime boundary. We need to point out that the radial coordinate is not Fourier transformed; hence if we want to construct correlation functions, we need to integrate the radial coordinate. Previous literature \cite{Armstrong:2022jsa} has developed the recursions for BG currents in (A)dS by the classical equation of motion. We briefly review a bit about it and apply the recursions here to derive the unifying relations \eqref{cor} in (A)dS. One can also find other perspectives to understand the BG currents in curved spacetime \cite{Herderschee:2022ntr,Cheung:2022pdk}. This section will mainly follow the notations in \cite{Armstrong:2022jsa}.

For convenience, we consider the $\text{AdS}_{d+1}$ in the Poincar\'{e} patch and the metric can be parameterized as
\begin{equation}\label{eq:adsmetric}
    ds^2=\frac{l^2}{z^2}(dz^2+\eta_{\mu\nu}dx^{\mu}dx^{\nu}),
\end{equation}
where $l$ is the radius and $0<z<\infty$. Here $\eta_{\mu\nu}$ is the flat boundary metric with Lorentzian signature and $\mu,\nu=0,1,\cdots,d-1$. It should be noted that the $\text{dS}_{d+1}$ metric can be derived from \eqref{eq:adsmetric} through analytical continuation $z\to -i\eta$ and $l\to -il$ after taking the boundary metric to be Euclidean. The equation of motion for Yang-Mills theory in AdS background is
\begin{equation}\label{eq:eomYMads}
g^{np}\partial_p\mb{F}_{mn}=ig^{np}[\mb{A}_p,\mb{F}_{mn}]+\mb{J}_m+g^{np}(\Gamma^q_{mp}\mb{F}_{qn}+\Gamma^q_{np}\mb{F}_{mq}),
\end{equation}
where $\mb{F}_{mn}=\partial_m\mb{A}_n-\partial_n\mb{A}_m-i[\mb{A}_m,\mb{A}_n]$ is the field strength, $\mb{A}_m$ is the Lie algebra valued for some gauge group with the generator $T^a$ and $\Gamma^p_{mn}=\frac{1}{2}g^{pq}(\partial_m g_{np}+\partial_n g_{mp}-\partial_pg_{mn})$ is the Christoffel symbols. And $\mb{J}_m$ is the current coupled to other fields and is the model dependent. 

The ansatz for multi-particle solutions of the Yang-Mills theory in AdS can be written as
\ie\label{eq:multiansatz}
\mb{A}_{\mu}(x,z)&=\frac{l}{z}\sum_I \ma{A}_{I\mu}(z)T^{a_I}e^{ik_I\cdot x}\\
\mb{A}_z(x,z)&=\frac{l}{z}\sum_I \alpha_I(z)T^{a_I}e^{ik_I\cdot x}\\
\mb{J}_m(x,z)&=\frac{l}{z}\sum_I \ma{J}_{Im}(z)T^{a_I}e^{ik_I\cdot x},
\fe
where $I$ is a bunch of letters $I=i_1\cdots i_s$ and each $i_{s}$ represents for a single particle state. Here $k_{I\mu}=k_{1\mu}+\cdots+k_{s\mu}$ is the total momentum for multi-particle states and $T^{a_I}=T^{a_{i_1}}\cdots T^{a_{i_s}}$. Now we let $U\cdot V=\eta_{\mu\nu}U^{\mu}V^{\nu}$. Substituting the ansatz \eqref{eq:multiansatz} into the equation of motion \eqref{eq:eomYMads}, we can obtain the multi-particle recursions. As we have mentioned above that we do not Fourier transferred bulk component to momentum space, then we write down the BG currents bulk and boundary components separately. The boundary components then are:
\begin{equation}\label{eq:gluonrecur}
	\begin{split}
		\frac{1}{z^2}(\mathcal{D}_I^2+d-1)\mb{\mathcal{A}}_{I\mu}&=ik_{I\mu}(\partial_z+\frac{2-d}{z})\alpha_I-\frac{l}{z}\mathcal{J}_{I\mu}+\frac{l}{z}\sum_{I=JK}\{(k_{K\mu}\alpha_K+2i\partial_z\mb{\mathcal{A}}_{K\mu})\alpha_J\\
		&+k_{K\mu}(\mb{\mathcal{A}}_J\cdot\mb{\mathcal{A}}_K)+\mb{\mathcal{A}}_{K\mu}[i(\partial_z-\frac{d}{z})\alpha_J-2(k_K\cdot\mb{\mathcal{A}}_J)]-(J\leftrightarrow K)\}\\
		&+\frac{l^2}{z^2}\sum_{I=JKL}\{[\alpha_J\alpha_K\ma{A}_{L\mu}+(\ma{A}_J\cdot\ma{A}_K)\ma{A}_{L\mu}-(K\leftrightarrow L)]\\
		&+[\alpha_K\alpha_L\ma{A}_{J\mu}+(\ma{A}_K\cdot\ma{A}_L)\ma{A}_{J\mu}-(J\leftrightarrow K)]\}.
	\end{split}	
\end{equation}
We can find that the boundary components $\ma{A}_{I\mu}$ are mixed with the bulk components $\alpha_I$. In other words, the higher points boundary terms of BG currents are contributed from not only the boundary but also the bulk components at lower points. And we can find that this is true for the higher points bulk components of BG currents:
\begin{equation}
	\begin{split}
		k_I^2\alpha_I=&\frac{l}{z}\sum_{I=JK}[2\alpha_K(k_K\cdot \mb{\mathcal{A}}_J)-2\alpha_J(k_J\cdot \mb{\mathcal{A}}_K)+i(\mb{\mathcal{A}}_J\cdot\partial_z\mb{\mathcal{A}}_K)-i(\mb{\mathcal{A}}_K\cdot\partial_z\mb{\mathcal{A}}_J)]\\
		&+\frac{l}{z}\mathcal{J}_{Iz}+\frac{l^2}{z^2}\sum_{I=JKL}[\alpha_K(\mb{\mathcal{A}}_J\cdot\mb{\mathcal{A}}_L)-\alpha_L(\mb{\mathcal{A}}_J\cdot\mb{\mathcal{A}}_K)+(J\leftrightarrow L)].
	\end{split}
\end{equation}
We should note that to construct the full BG currents or the correlation function, we should sum over all the possible deconcatenation. As we have explained in previous discussion Fig. \ref{fig:binarytree}, each deconcatenation represents a specific binary tree. Thus, if we focus on some specific binary tree, we can only pick up the corresponding binary tree and ignore the deconcatenation sum. The starting point of our recursion is the single point currents which are imposed as $\mathcal{A}_{i\mu}=\phi_{i}(z)\epsilon_{i\mu}$ and $\alpha_{i}=0$ separately for boundary and bulk components. Here $\ma{D}_I^2=\ma{D}_{k_I}^2$ is the d'Alembert operator in AdS:
\begin{equation}
    \ma{D}_{k_I}^2=z^2\partial^2_z+(1-d)z\partial_z-z^2k_I^2,
\end{equation}
and the signle point current $\phi_{i}(z)$ satisfies the Klein-Gordon equation $(\ma{D}_{i}^2-M^{2})\phi_{i}(z)=0$. We have also imposed the boundary transversal gauge: $\eta^{\mu\nu}\partial_{\mu}A_{\nu}=0$. The information of 4-pt vertices is now included in the terms with the sum over $I=JKL$, which is equivalent to the double sum we have mentioned before since we have summed over all the possible binary trees. By the way, we should emphasize that gluons must be massless; however, the mass we consider here is the effective mass that comes from we take \eqref{eq:multiansatz} to get a locally flat patch. 

The recursions for scalars in the adjoint representation of the gauge group can also be derived from what we have done above. Similarly, we can write down the ansatz for the multi-particle solution to the scalars equation of motion:
\begin{equation}
    \phi=\sum_I \Phi_I(z)T^{a_I}e^{ik_I\cdot x}
\end{equation}
As we have said, the interaction between scalar and gauge field is through the current $\ma{J}$ presented in \eqref{eq:multiansatz}. Generally speaking, the interaction should be model dependent. However, we focus on the minimal coupled interaction between scalars and gauge fields for simplicity. In this case the current should be $\mb{J}_m=[(i\partial_m\phi+[\mb{A}_m,\phi]),\phi]$ in \eqref{eq:eomYMads}. Specifically, the minimal coupled interaction terms in BG currents can be written as
\ie\label{cur}
\mathcal{J}_{I\mu}&=\sum_{I=JK}(-k_{J\mu}\Phi_{J}\Phi_{K}+k_{K\mu}\Phi_{K}\Phi_{J})\\
&+\frac{l}{z}\sum_{I=JKL}[\tilde{\mathcal{A}}_{J\mu}\Phi_{K}\Phi_{L}-\tilde{\mathcal{A}}_{K\mu}\Phi_{J}\Phi_{L}-\Phi_{J}\tilde{\mathcal{A}}_{K\mu}\Phi_{L}+\Phi_{J}\tilde{\mathcal{A}}_{L\mu}\Phi_{K}]\\
\mathcal{J}_{Iz}&=\sum_{I=JK}(-i\Phi_{J}\overleftrightarrow{\partial_{z}}\Phi_{K})+\frac{l}{z}\sum_{I=JKL}[\tilde{\alpha}_{J}\Phi_{K}\Phi_{L}-\tilde{\alpha}_{K}\Phi_{J}\Phi_{L}-\Phi_{J}\tilde{\alpha}_{K}\Phi_{L}+\Phi_{J}\tilde{\alpha}_{K}\Phi_{L}],
\fe
where the tilde is used to distinguish the pure gluon theory and the scalar theory we discussed above.

Similarly, we can also impose the boundary transversal gauge. Consequently, we can obtain the recursions for scalars minimally coupled to gauge fields:
\begin{equation}\label{eq:scalarrecur}
    \begin{split}
        \frac{1}{z^2}(\ma{D}_I^2-M^2)\Phi_I&=\frac{l}{z}\sum_{I=JK}[2\Phi_J(k_J\cdot\tilde{\ma{A}}_K)-i(\Phi_J\partial_z\tilde{\alpha}_K+2\tilde{\alpha}_K\partial_z\Phi_J-\frac{d}{z}\Phi_J\tilde{\alpha}_K)-(J\leftrightarrow K)]\\
        &+\frac{l^2}{z^2}\sum_{I=JKL}[(\tilde{\ma{A}}_J\cdot\tilde{\ma{A}}_K)\Phi_L-(\tilde{\ma{A}}_J\cdot\tilde{\ma{A}}_L)\Phi_K+\tilde{\alpha}_J\tilde{\alpha}_K\Phi_L-\tilde{\alpha}_J\tilde{\alpha}_L\Phi_K+(J\leftrightarrow L)].
    \end{split}
\end{equation}

Note that for the currents above, the flat limit can be realized by taking $\mathcal{A}_{I\mu}(\tilde{\mathcal{A}}_{I\mu},\mathcal{J}_{Im})\to \frac{z}{l}\mathcal{A}_{I\mu}(\frac{z}{l}\tilde{\mathcal{A}}_{I\mu},\frac{z}{l}\ma{J}_{Im})$ and $\phi_{i}(z)(\Phi_{i}(z))=1$, then taking the limit $l/z\to 1$. Then $\alpha_{I}(\tilde{\alpha}_{I})$ will be zero, and the propagators and the ansatz \eqref{eq:multiansatz} will be the flat ones. This process can be understood as follows: first restore $z$ (or not) in the ansatz according to \eqref{eq:multiansatz}, then take the mode of flat Klein-Gordon equation to construct one-particle states and the metric to be flat. After we take this limit, the currents in (A)dS coincide with the flat currents.

As in the flat case, we can also construct tree-level $n$-pt correlation functions from $(n-1)$-pt BG currents in (A)dS. We define them for gluons as
\ie\label{eq:ymcorrelation}
A_{\rm YM\,}(1,2,\cdots,N)=-\frac{1}{N}\int\frac{dz}{z^{d+1}}\mathcal{A}_{N}\cdot(\ma{D}_{1\cdots N-1}^2+d-1)\ma{A}_{1\cdots N-1}+\text{cyclic}(1,2,\cdots,N),
\fe
and for scalars as
\ie\label{scacor}
A_{\rm S\,}(1,2,\cdots,N)=-\frac{1}{N}\int\frac{dz}{z^{d+1}}\Phi_{N}(\ma{D}_{1\cdots N-1}^2-M^{2})\Phi_{1\cdots N-1}+\text{cyclic}(1,2,\cdots,N).
\fe

Starting with the equation of motion for minimal coupling scalars and gauge fields, we have written down the recursion relations for both scalars and gauge fields in (A)dS background. In the following sections, we use these recursions to derive a type of unifying relations \eqref{cor} in (A)dS \ref{sec:unifyingads} and calculate some specific cosmological correlators which are related to the cosmic observations \ref{sec:applications}.

\section{A Type of Unifying Relations in (A)dS}
\label{sec:unifyingads}
In section \ref{sec:unifyingflat}, we have reviewed some of the unifying relations \eqref{uni}, \eqref{uni2} and \eqref{cor} in flat spacetime. As a natural generalization, it is desirable to figure out whether the unifying relations \eqref{uni}, \eqref{uni2} and \eqref{cor} are valid in curved spacetime. For simplicity, we only consider the simplest case \eqref{cor} in this work and put the other relations \eqref{uni} and \eqref{uni2} in future work.  In this section \ref{sec:unifyingads}, we use the BG currents recursions in (A)dS to bridge the correlation functions between pure YM theory and scalar theory minimal coupled with gluons by acting some trace operators $\ma{T}[ij]$ on YM side. To be familiar with the unifying relations \eqref{cor} we aim to prove, we take the four points \ref{subsec:4points} and six points \ref{subsec:6points} correlation functions as some examples. Furthermore, we can find the validity of unifying relations \eqref{cor} in such examples. In subsection \ref{subsec:proof}, we present a rigorous proof to unifying relations \eqref{cor} in (A)dS. In the final subsection, we discuss a more helpful type of unifying relations: the generalization of the relation in the subsection \ref{subsec:proof}. In this section, we will focus on the relations between an arbitrary binary tree on both sides, and we use the following notation to denote the deconcatenation: $I=JK, J=AB, K=CD$ for a given binary tree. We will not consider mixed currents of scalars and gluons until subsection \ref{more}. In this section, the currents with a Lie monomial $\Gamma$ as an upper label denote the currents for a given binary tree corresponds to this $\Gamma$, and after summing over all possible $\Gamma$ we will get the total BG currents we introduced before.  The recursion for the currents with a $\Gamma$ is limited by the given binary tree so that each lower point current should bring a sub-tree which is consistent with its boundary legs as its upper label. For more explainment, see \cite{Frost:2020eoa}. Note that for one-particle states or 2-pt BG currents, the currents with a $\Gamma$ are the same as the total BG currents, i.e., for $\Gamma=i$, $\Phi_{i}^{i}=\Phi_{i}$, and for $\Gamma=[i,j]$, $\Phi_{ij}^{[i,j]}=\Phi_{ij}$, because there is only one possible binary tree for one-particle state currents or 2-pt currents. For higher point currents, things are different. For example, for 3-pt currents, we have $\Phi_{123}=\Phi^{[[1,2],3]}_{123}+\Phi^{[1,[2,3]]}_{123}$, which means that for the 3-pt currents, there are two different traces and can be diagrammatically represented by two binary trees Fig. \ref{fig:binarytree}.

Before our proof, there are some comments we need to emphasize again. The scalar theory with a minimal coupling with gluons can be understood as the second color group of the YM scalar being the $U(1)$ group. Compared with the general YMS theory, minimal coupling scalar theory has only one color index for scalars; hence it only has single color-ordered correlation functions. Moreover, it has no $\phi^{3}$ and $\phi^{4}$ vertices. In this case, scalars with no gluon propagator between them are called ``in the same trace". Each trace can only have two scalars because of the lack of $\phi^{3}$ and $\phi^{4}$ vertices. Then we only need to let some $\mathcal{T}[ij]$ operators act on the YM currents and prove that we will obtain the corresponding scalar currents. This section assumes that the spacetime background is AdS, while the analysis is also valid for the dS case. For convenience, we take $l=1$, and the scalars are conformal coupling, i.e., $M^{2}=1-d$, to make the propagators of gluons and scalars the same. In fact, conformally coupling scalars are also massless. The effective mass $M$ comes from the curved spacetime.

The relations we want to prove are
\ie\label{ppp}
\mathcal{T}^{Xn}(\mathcal{D}^2_{I}+d-1)\mathcal{A}_{I}\cdot\epsilon_{n}&=(-1)^{\frac{|X|-1}{2}}(\mathcal{D}^2_{I}+d-1)\Phi_{I,X}\ \ (|X|\in \text{odd})\\
\mathcal{T}^{X}\mathcal{A}_{I}\cdot v&=(-1)^{\frac{|X|}{2}}\tilde{\mathcal{A}}_{I,X}\cdot v\ \ (|X|\in \text{even})\\
\mathcal{T}^{X}\alpha_{I}&=(-1)^{\frac{|X|}{2}}\tilde{\alpha}_{I,X}\ \ (|X|\in \text{even}),
\fe
where $\ma{T}^{X}$ is the operator appeared in \eqref{cor}, and $\Phi_{I,X}$ means that the boundary legs of the scalar current $\Phi_{I}$ in the word $X\subseteq I$ (if $X$ is not a subset of $I$ we will get $\Phi_{I,X}=0$) are scalars and others are gluons (similar for the gluon currents $\tilde{\mathcal{A}}_{I,X\mu}$ and $\tilde{\alpha}_{I,X}$). For example, $\Phi_{123,12}$ means that the scalar current $\Phi_{123}$ has 2 scalar boundary legs 1 and 2, and the leg 3 is a gluon boundary leg. The vector $v_{\mu}$ here can be any vector, such as a momentum vector or a polarization vector. We will prove the case $X=I$ first and then the general case. For simplicity, if $X=I$, we will omit the label $X$ of the currents in the scalar theory, for example, for scalar currents $\Phi_{I,I}:=\Phi_{I}$. These relations \eqref{ppp} are the generalization of the unifying relation \eqref{cor} for BG currents with different bulk legs in (A)dS; hence we also call these relations ``unifying relations" for BG currents. After we construct correlation functions from these BG currents, we will reproduce \eqref{cor} in (A)dS case. In this section, we will prove a stronger statement where the unifying relation is valid even for each given trace or deconcatenation $\Gamma$:
\ie\label{pppp}
\mathcal{T}^{Xn}(\mathcal{D}^2_{I}+d-1)\mathcal{A}^{\Gamma}_{I}\cdot\epsilon_{n}&=(-1)^{\frac{|X|-1}{2}}(\mathcal{D}^2_{I}+d-1)\Phi_{I,X}^{\Gamma}\ \ (|X|\in \text{odd})\\
\mathcal{T}^{X}\mathcal{A}^{\Gamma}_{I}\cdot v&=(-1)^{\frac{|X|}{2}}\tilde{\mathcal{A}}^{\Gamma}_{I.X}\cdot v\ \ (|X|\in \text{even})\\
\mathcal{T}^{X}\alpha^{\Gamma}_{I}&=(-1)^{\frac{|X|}{2}}\tilde{\alpha}^{\Gamma}_{I,X}\ \ (|X|\in \text{even}).
\fe
After summing over currents for all possible binary trees consistent with $I$, i.e. by using the relation
\ie
J=\sum_{\Gamma}J^{\Gamma},
\fe
where $J$ can be any type of BG currents (we omit the possible vector index $\mu$), we will come back to \eqref{ppp}. One may also confused with the factor of some minus signs, and we will explain this important thing later. By the way, we still use the boundary Lorenz gauge in this section. This is valid because what we want to prove is the relation of the correlation functions \eqref{cor} in (A)dS case which is independent of the gauge choice, and it is enough to prove that the relation for currents \eqref{pppp} holds in the boundary Lorenz gauge.

\subsection{4-points Unifying Relations in (A)dS}
\label{subsec:4points}
In this subsection and the next subsection we only consider the case that all legs on boundary are the same particles, which corresponds to $X=I$ in the equation \eqref{pppp}. To convince the validity of unifying relations \eqref{cor} in AdS, let us consider 4-pt correlation functions first, which correspond to 3-pt BG currents. Without loss of generality, we only consider the following binary tree in this subsection (FIG. \ref{fig:4points}). In this case, the binary tree corresponds to the following deconcatenation: $I=123, J=12, K=3, A=1, B=2$ and there are no contributions from $C$ or $D$ terms. The corresponding Lie monomial of this binary tree is [[1,2],3]. Thus here we have $\Gamma=[[1,2],3]$, and its proper sub-trees are [1,2], 1, 2 and 3. For this binary tree, the only trace structure of the corresponding scalar correlator is $(12|34)$, and the corresponding operator is $\mathcal{T}[12]\mathcal{T}[34]$. By using the recursions \eqref{eq:gluonrecur}, it is not hard to find the gluon currents:
\ie\label{eq:3ptgluoncurrent}
\frac{1}{z^{2}}(\mathcal{D}_{123}^2+d-1)\mathcal{A}^{[[1,2],3]}_{123\mu}&=ik_{123\mu}(\partial_{z}+\frac{2-d}{z})\alpha^{[[1,2],3]}_{123}+\frac{1}{z}[2i\partial_{z}\mathcal{A}^{3}_{3\mu}\alpha^{[1,2]}_{12}+k_{3\mu}(\mathcal{A}^{[1,2]}_{12}\cdot\mathcal{A}^{3}_{3})\\
&+\mathcal{A}^{3}_{3\mu}(i(\partial_{z}-\frac{d}{z})\alpha^{[1,2]}_{12}-2(k_{3}\cdot\mathcal{A}^{[1,2]}_{12}))-k_{12\mu}(\mathcal{A}^{3}_{3}\cdot\mathcal{A}^{[1,2]}_{12})+2\mathcal{A}^{[1,2]}_{12\mu}(k_{12}\cdot\mathcal{A}^{3}_{3})]\\
&+\frac{1}{z^{2}}[(\mathcal{A}^{2}_{2}\cdot\mathcal{A}^{3}_{3})\mathcal{A}^{1}_{1\mu}-(\mathcal{A}^{1}_{1}\cdot\mathcal{A}^{3}_{3})\mathcal{A}^{2}_{2\mu}],
\fe
where $\ma{A}^{[1,2]}_{12}$ and $\alpha^{[1,2]}_{12}$ are the boundary and $z$ components for gluon 2-pt current, and these 2-pt currents can also be solved out from the recursions relation \eqref{eq:gluonrecur}: 
\ie\label{eq:2ptgluoncurrent}
\frac{1}{z^{2}}(\mathcal{D}_{12}^2+d-1)\mathcal{A}^{[1,2]}_{12\mu}&=ik_{12\mu}(\partial_{z}+\frac{2-d}{z})\alpha^{[1,2]}_{12}+\frac{1}{z}[k_{2\mu}(\mathcal{A}^{1}_{1}\cdot\mathcal{A}^{2}_{2})-2\mathcal{A}^{2}_{2\mu}(k_{2}\cdot\mathcal{A}^{1}_{1})-(1\leftrightarrow2)]\\
\alpha^{[1,2]}_{12}&=\frac{1}{zk_{12}^{2}}i(\epsilon_{1}\cdot\epsilon_{2})\phi_{1}\overleftrightarrow{\partial} \phi_{2}
\fe
Here we have denoted that $u\overleftrightarrow{\partial} v=u\partial_z v-v\partial_z u$ for simplicity. To show the relations we want to prove, we can act the operator $\mathcal{T}[12]\mathcal{T}[34]$ on the 3-pt current \eqref{eq:3ptgluoncurrent} directly. This operation will kill the terms that do not contain $\epsilon_3\cdot\epsilon_4$. Then the remaining terms after substituting 2-pt currents \eqref{eq:2ptgluoncurrent} are (from now on we use $\sim$ to indicate that what we write down are the ``remaining terms"):
\ie\label{eq:3ptgluoncurrent1}
\frac{1}{z^{2}}(\mathcal{D}_{123}^2+d-1)\mathcal{A}^{[[1,2],3]}_{123\mu}&\sim\frac{1}{z}[2i\partial_{z}\mathcal{A}^{3}_{3\mu}\alpha^{[1,2]}_{12}+\mathcal{A}^{3}_{3\mu}(i(\partial_{z}-\frac{d}{z})\alpha^{[1,2]}_{12}-2(k_{3}\cdot\mathcal{A}^{[1,2]}_{12}))]\\
&=\frac{1}{z}[-2\epsilon_{3\mu}\partial_{z}\phi_{3}(\frac{1}{zk^{2}_{12}}(\epsilon_{1}\cdot\epsilon_{2})\phi_{1}\overleftrightarrow{\partial}\phi_{2})\\
&-\epsilon_{3\mu}\phi_{3}\{(\partial_{z}-\frac{d}{z})(\frac{1}{zk^{2}_{12}}(\epsilon_{1}\cdot\epsilon_{2})\phi_{1}\overleftrightarrow{\partial}\phi_{2})\\
&+2\frac{1}{\mathcal{D}^2_{12}+d-1}[-k_{3}\cdot k_{12}(\partial_{z}-\frac{d-2}{z})(\frac{1}{zk^{2}_{12}}(\epsilon_{1}\cdot\epsilon_{2})\phi_{1}\overleftrightarrow{\partial}\phi_{2})\\
&+\frac{1}{z}k_{3}\cdot(k_{2}-k_{1})(\epsilon_{1}\cdot\epsilon_{2})\phi_{1}\phi_{2}]\}].
\fe
The final result after we act $\ma{T}^{1234}$ (equals to the result we act $\mathcal{T}[12]\mathcal{T}[34]$ since the sum of the contribution of the others is 0) on the 3-pt current \eqref{eq:3ptgluoncurrent1} is 
\ie
\mathcal{T}^{1234}(\mathcal{D}^2_{123}+d-1)\mathcal{A}^{[[1,2],3]}_{123}\cdot\epsilon_{4}&=\mathcal{T}[12]\mathcal{T}[34](\mathcal{D}^2_{123}+d-1)\mathcal{A}^{[[1,2],3]}_{123}\cdot\epsilon_{4}\\
&=z[-2\partial_{z}\phi_{3}(\frac{1}{zk^{2}_{12}}\phi_{1}\overleftrightarrow{\partial}\phi_{2})-\phi_{3}(\partial_{z}-\frac{d}{z})(\frac{1}{zk^{2}_{12}}\phi_{1}\overleftrightarrow{\partial}\phi_{2})\\
&+2\phi_{3}\frac{1}{\mathcal{D}^2_{12}+d-1}[k_{3}\cdot k_{12}z^{2}(\partial_{z}-\frac{d-2}{z})(\frac{1}{zk^{2}_{12}}\phi_{1}\overleftrightarrow{\partial}\phi_{2})\\
&-zk_{3}\cdot(k_{2}-k_{1})\phi_{1}\phi_{2}]].
\fe
\begin{figure}[H]
\centering
\begin{tikzpicture}[line width=1pt,scale=1.5]
\draw[scalarnoarrow](0.5,0)--(3.5,0);
\draw[fermionnoarrow](2,-1)--(3,0);
\draw[fermionnoarrow](1.5,-0.5)--(2,0);
\draw[fermionnoarrow](1.5,-0.5)--(1,0);
\draw[fermionnoarrow](2,-1)--(1.5,-0.5);
\draw[fermionnoarrow](2,-1)--(2,-1.5);
\draw[fermionnoarrow,fill=pink] (1.5,-0.5) circle (.05cm);
\draw[fermionnoarrow,fill=pink] (2,-1) circle (.05cm);
\node at (1,0.2) {1};
\node at (2,0.2) {2};
\node at (3,0.2) {3};
\end{tikzpicture}
\caption{Binary tree [[1,2],3] for 3-point currents. One can understand the dashed line as the boundary of spacetime. In our following calculation, we focus on the AdS background, but this representation is also correct for the dS background.}
\label{fig:4points}
\end{figure}

For scalar currents, there is no gluon external leg hence $\tilde{\ma{A}}_{i}^{\mu}=\tilde{\ma{A}}_{i}^{\Gamma\mu}=0$. We can write down the 2-pt currents first by the recursions \eqref{cur} and \eqref{eq:scalarrecur}:
\ie\label{eq:2ptscalarcurrent}
\mathcal{J}^{[1,2]}_{12\mu}&=(k_{2\mu}-k_{1\mu})\Phi^{1}_{1}\Phi^{2}_{2}\\
\mathcal{J}^{[1,2]}_{12z}&=-i\Phi^{1}_{1}\overleftrightarrow{\partial}\Phi^{2}_{2}\\
\tilde{\alpha}^{[1,2]}_{12}&=\frac{-i}{zk_{12}^{2}}(\Phi^{1}_{1}\overleftrightarrow{\partial} \Phi^{2}_{2})\\
\frac{1}{z^{2}}(\mathcal{D}_{12}^2+d-1)\tilde{\mathcal{A}}^{[1,2]}_{12\mu}&=ik_{12\mu}(\partial_{z}+\frac{2-d}{z})\tilde{\alpha}^{[1,2]}_{12}-\frac{1}{z}(k_{2\mu}-k_{1\mu})\Phi^{1}_{1}\Phi^{2}_{2}
\fe
Apply the recursions \eqref{eq:scalarrecur} for scalar into 3-pt scalar correlation and take the 2-pt current \eqref{eq:2ptscalarcurrent} into consideration, we can obtain the 3-pt correlation function for the conformally coupled scalar:
\ie
(\mathcal{D}_{123}^2+d-1)\Phi^{[[1,2],3]}_{123}=&-2(\frac{k_{12}\cdot k_{3}}{k_{12}^{2}}(k_{2}^{2}-k_{1}^{2})-(k_{2}-k_{1})\cdot k_{3})z\Phi^{3}_{3}(\frac{1}{\mathcal{D}^2_{12}+d-1}z\Phi^{1}_{1}\Phi^{2}_{2})\\
&+\frac{1}{k_{12}^{2}}((k^{2}_{2}-k_{1}^{2})\Phi^{1}_{1}\Phi^{2}_{2}\Phi^{3}_{3}+2\Phi^{1}_{1}\overleftrightarrow{\partial} \Phi^{2}_{2}\partial_{z}\Phi^{3}_{3}-2\Phi^{3}_{3}\frac{\Phi^{1}_{1}\overleftrightarrow{\partial} \Phi^{2}_{2}}{z}).
\fe

After some derivation, we find that if we let $\Phi_{i} (\Phi^{i}_{i})=\phi_{i}$ (From now on, we always take this condition. It is reasonable because the Klein-Gordon equation of both theories are the same.), we will have
\ie
\mathcal{T}^{1234}(\mathcal{D}_{123}^2+d-1)\mathcal{A}^{[[1,2],3]}_{123}\cdot\epsilon_{4}=-(\mathcal{D}_{123}^2+d-1)\Phi^{[[1,2],3]}_{123}.
\fe

The other two relations in \eqref{pppp} for this binary tree and its sub-trees are manifest during our calculation. Note that some other trace structures are included in the gluon current above. These terms, which come from the 4-pt vertices of gluons' action, will lead to an extra 4-pt scalar contact terms after acting, say, the operator $\ma{T}[13]\ma{T}[24]$. These terms are the same as the flat case \cite{Cheung:2017ems} and will be canceled here after summing all trace structures. In the higher point cases, these terms can be realized by the vertices of two scalars and two gluons.

\subsection{6-points Unifying Relations in (A)dS}
\label{subsec:6points}
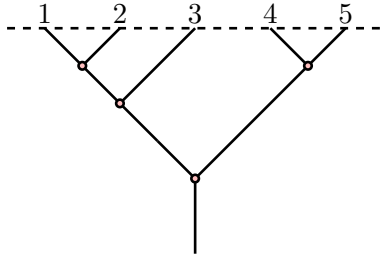
\begin{figure}[H]
\centering
\begin{tikzpicture}[line width=1pt,scale=1]
\draw[scalarnoarrow](-0.5,0)--(4.5,0);
\draw[fermionnoarrow](2,-2)--(4,0);
\draw[fermionnoarrow](2,-2)--(1.5,-1.5);
\draw[fermionnoarrow](1.5,-1.5)--(1,-1);
\draw[fermionnoarrow](1,-1)--(0.5,-0.5);
\draw[fermionnoarrow](0.5,-0.5)--(0,0);
\draw[fermionnoarrow](0.5,-0.5)--(1,0);
\draw[fermionnoarrow](1,-1)--(2,0);
\draw[fermionnoarrow](3.5,-0.5)--(3,0);
\draw[fermionnoarrow](2,-2)--(2,-3);
\draw[fermionnoarrow,fill=pink] (0.5,-0.5) circle (.05cm);
\draw[fermionnoarrow,fill=pink] (1,-1) circle (.05cm);
\draw[fermionnoarrow,fill=pink] (3.5,-0.5) circle (.05cm);
\draw[fermionnoarrow,fill=pink] (2,-2) circle (.05cm);
\node at (0,0.2) {1};
\node at (1,0.2) {2};
\node at (2,0.2) {3};
\node at (3,0.2) {4};
\node at (4,0.2) {5};
\end{tikzpicture}
\caption{Binary tree [[[1,2],3],[4,5]] for 5-points currents. One can understand the dashed line as the boundary of spacetime. We focus on the AdS background in our following calculation, but this representation is also correct for the dS background.}
\label{fig:6points}
\end{figure}

Now we turn to the 6-pt correlation functions and consider the 5-pt currents. In this subsection, we choose the binary tree in (FIG. \ref{fig:6points}), and the trace structure of the corresponding scalar correlator is $(12|36|45)$. This time the deconcatenation is $J=123, K=45, A=12, B=3, C=4, D=5$, and the corresponding Lie monomial is [[[1,2],3],[4,5]]. Thus we have $
\Gamma=[[[1,2],3],[4,5]]$, and the proper sub-trees are [[1,2],3], [4,5], [1,2], 1, 2, 3, 4, 5. Since the binary tree of the 3-pt currents we have computed in the previous subsection is one of the sub-trees of this 5-pt current case, we can use the results in the previous subsection to evaluate the recursion.

We can also write down the 5-pt scalar current from the scalar recursions \eqref{eq:scalarrecur}. 
\ie
(\mathcal{D}_{12345}^2+d-1)\Phi^{[[[1,2],3],[4,5]]}_{12345}=&z[2\Phi^{[[1,2]],3]}_{123}(k_{123}\cdot\tilde{\mathcal{A}}^{[4,5]}_{45})\\
&-i(\Phi^{[[1,2],3]}_{123}\partial_{z}\tilde{\alpha}^{[4,5]}_{45}+2\tilde{\alpha}^{[4,5]}_{45}\partial_{z}\Phi^{[[1,2],3]}_{123}-\frac{d}{z}\Phi^{[[1,2],3]}_{123}\tilde{\alpha}^{[4,5]}_{45})]\\
&-[(\tilde{\mathcal{A}}^{[4,5]}_{45}\cdot\tilde{\mathcal{A}}^{[1,2]}_{12})\Phi^{3}_{3}+\tilde{\alpha}^{[1,2]}_{12}\tilde{\alpha}^{[4,5]}_{45}\Phi^{3}_{3}].
\fe

For gluon currents, note that some terms have no contribution after acting the unifying operator $\ma{T}[12]\ma{T}[45]\ma{T}[36]$ (for the same reason as 3-pt currents, $\ma{T}^{123456}$ is equivalent to $\ma{T}[12]\ma{T}[45]\ma{T}[36]$). The terms which will be non-vanishing  after acting $\ma{T}[36]$ are
\ie\label{eq:5ptscalarcurrent}
(\mathcal{D}_{12345}^2+d-1)\mathcal{A}^{[[[1,2],3],[4,5]]}_{12345\mu}\sim&-z[2i\partial_{z}\mathcal{A}^{[[1,2],3]}_{123\mu}\alpha^{[4,5]}_{45}+\mathcal{A}^{[[1,2],3]}_{123\mu}(i(\partial_{z}-\frac{d}{z})\alpha^{[4,5]}_{45}-2(k_{123}\cdot\mathcal{A}^{[4,5]}_{45}))]\\
&-[(\mathcal{A}^{[1,2]}_{12}\cdot\mathcal{A}^{[4,5]}_{45})\mathcal{A}^{3}_{3\mu}-\alpha^{[1,2]}_{12}\alpha^{[4,5]}_{45}\mathcal{A}^{3}_{3\mu}].
\fe
After substituting lower point currents, one can easily find that
\ie\label{eq:5ptgluoncurrent}
\ma{T}^{123456}(\mathcal{D}_{12345}^2+d-1)\mathcal{A}^{[[[1,2],3],[4,5]]}_{12345}\cdot\epsilon_{6}=(\mathcal{D}_{12345}^2+d-1)\Phi^{[[[1,2],3],[4,5]]}_{12345}.
\fe
It is not hard to find the other two relations in \eqref{pppp} for this binary tree and its sub-trees are also correct. For other binary trees, things are the same. Hence it is natural to expect that \eqref{pppp} is correct for $X=I$ and for every binary tree.

\subsection{Proof of the Unifying Relations in (A)dS for $X=I$}
\label{subsec:proof}
Now we give the full proof for the relation \eqref{pppp} for the case $X=I$. In this section, we will prove that \eqref{pppp} actually holds for any given binary tree by induction. As before, all currents appear in this subsection are just the currents for a given binary tree, and assume that we have the deconcatenation $I=JK, J=AB, K=CD$ for this given binary tree. The Lie monomials correspond to these words are denoted by $\Gamma_{J}$, $\Gamma_{K}$ and so on, as sub-trees of $\Gamma=\Gamma_{I}$.

The relation we want to prove in this subsection is:
\ie
\label{proof}
&\mathcal{T}^{In}\mathcal{A}^{\Gamma}_{I}\cdot\epsilon_{n}=(-1)^{\frac{|I|-1}{2}}\Phi^{\Gamma}_{I}\ \ (|I|\ \text{odd})\\
&\mathcal{T}^{I}\mathcal{A}^{\Gamma}_{I}\cdot v=(-1)^{\frac{|I|}{2}}\tilde{\mathcal{A}}^{\Gamma}_{I}\cdot v\ \ (|I|\ \text{even}) \\
&\mathcal{T}^{I}\alpha^{\Gamma}_{I}=(-1)^{\frac{|I|}{2}}\tilde{\alpha}^{\Gamma}_{I}\ \ (|I|\ \text{even}).
\fe
These are still the relations between given binary tree and sum over all the possible binary trees, we can obtain the relations for the full currents. The gluon currents for a given binary tree $\Gamma$ can be written as follows:
\ie\label{eq:recurgiventree}
\frac{1}{z^{2}}(\mathcal{D}^2_{I}+d-1)\mathcal{A}^{\Gamma}_{I}\cdot\epsilon_{n}&=ik_{I}\cdot\epsilon_{n}[\partial_{z}+(2-d)/z]\alpha^{\Gamma}_{I}+\frac{1}{z}[(k_{K}\cdot\epsilon_{n}\alpha^{\Gamma_{K}}_{K}+2i\partial_{z}\mathcal{A}^{\Gamma_{K}}_{K}\cdot\epsilon_{n})\alpha^{\Gamma_{J}}_{J}\\
&~~~+k_{K}\cdot\epsilon_{n}(\mathcal{A}^{\Gamma_{J}}_{J}\cdot\mathcal{A}^{\Gamma_{K}}_{K})+\mathcal{A}^{\Gamma_K}_{K}\cdot\epsilon_{n}[i(\partial_{z}-d/z)\alpha^{\Gamma_J}_{J}-2k_{K}\cdot\mathcal{A}^{\Gamma_J}_{J}]-(J\leftrightarrow K)]\\
&~~~+\frac{1}{z^{2}}[\alpha^{\Gamma_J}_{J}\alpha^{\Gamma_C}_{C}\mathcal{A}^{\Gamma_D}_{D}\cdot\epsilon_{n}+(\mathcal{A}^{\Gamma_J}_{J}\cdot\mathcal{A}^{\Gamma_C}_{C})\mathcal{A}^{\Gamma_D}_{D}\cdot\epsilon_{n}-(C\leftrightarrow D)+\alpha^{\Gamma_K}_{K}\alpha^{\Gamma_B}_{B}\mathcal{A}^{\Gamma_A}_{A}\cdot\epsilon_{n}\\
&~~~+(\mathcal{A}^{\Gamma_B}_{B}\cdot\mathcal{A}^{\Gamma_K}_{K})\mathcal{A}^{\Gamma_A}_{A}\cdot\epsilon_{n}-(A\leftrightarrow B)]. ~~~(I=JK,J=AB,K=CD)
\fe
Note that here we have no deconcatenation sums because we only consider a certain binary tree and all words appear in the equation above are known by reading this given binary tree. Besides, we want to note that in the left-hand side of \eqref{eq:recurgiventree}, we have the word $I$ while in the right-hand side, we have word $J,K,C$ and $D$. These are related through the given deconcatenation: $I=JK,J=AB,K=CD$. We show the deconcatenation at the end of equations to avoid the ambiguities. And we add the deconcatenation indication in the following discussion where may bring us ambiguities. We can also act the trace operator $\mathcal{T}^{In}$ on the gauge field currents. From the experience of what we have learned from the 4-pt and 6-pt calculations, we know that there are only some effective terms, i.e., the terms will not be annihilated by $\ma{T}[in]$ for a certain letter $i$ in the word $I$. These terms are (we still use the notation $\sim$ here):
\ie
\frac{1}{z^{2}}(\mathcal{D}^2_{I}+d-1)\mathcal{A}^{\Gamma}_{I}\cdot\epsilon_{n}&\sim\frac{1}{z}[(2i\partial_{z}\mathcal{A}^{\Gamma_K}_{K}\cdot\epsilon_{n})\alpha^{\Gamma_J}_{J}+\mathcal{A}^{\Gamma_K}_{K}\cdot\epsilon_{n}[i(\partial_{z}-d/z)\alpha^{\Gamma_J}_{J}-2k_{K}\cdot\mathcal{A}^{\Gamma_J}_{J}]-(J\leftrightarrow K)]\\
&~~~+\frac{1}{z^{2}}[\alpha^{\Gamma_J}_{J}\alpha^{\Gamma_C}_{C}\mathcal{A}^{\Gamma_D}_{D}\cdot\epsilon_{n}+(\mathcal{A}^{\Gamma_J}_{J}\cdot\mathcal{A}^{\Gamma_C}_{C})\mathcal{A}^{\Gamma_D}_{D}\cdot\epsilon_{n}-(C\leftrightarrow D)+\alpha^{\Gamma_K}_{K}\alpha^{\Gamma_B}_{B}\mathcal{A}^{\Gamma_A}_{A}\cdot\epsilon_{n}\\
&~~~+(\mathcal{A}^{\Gamma_B}_{B}\cdot\mathcal{A}^{\Gamma_K}_{K})\mathcal{A}^{\Gamma_A}_{A}\cdot\epsilon_{n}-(A\leftrightarrow B)].~~~(I=JK,J=AB,K=CD)
\fe
The scalar currents can also be obtained from the scalar recursion \eqref{eq:scalarrecur}:
\ie
\frac{1}{z^{2}}(\mathcal{D}^2_{I}+d-1)\Phi^{\Gamma}_{I}&=\frac{1}{z}[-2\Phi^{\Gamma_K}_{K}(k_{K}\cdot\tilde{\mathcal{A}}^{\Gamma_J}_{J})+2i\tilde{\alpha}^{\alpha_J}_{J}\partial_{z}\Phi_{K}+i\Phi^{\Gamma_K}_{K}(\partial_{z}-d/z)\tilde{\alpha}^{\Gamma_J}_{J}-(J\leftrightarrow K)]\\
&~~~+\frac{1}{z^{2}}[(\tilde{\mathcal{A}}^{\Gamma_J}_{J}\cdot\tilde{\mathcal{A}}^{\Gamma_C}_{C})\Phi^{\Gamma_D}_{D}+\tilde{\alpha}^{\Gamma_J}_{J}\tilde{\alpha}^{\Gamma_C}_{C}\Phi^{\Gamma_D}_{D}-(C\leftrightarrow D)+(\tilde{\mathcal{A}}^{\Gamma_K}_{K}\cdot\tilde{\mathcal{A}}^{\Gamma_B}_{B})\Phi^{\Gamma_A}_{A}\\
&~~~+\tilde{\alpha}^{\Gamma_K}_{K}\tilde{\alpha}^{\Gamma_B}_{B}\Phi^{\Gamma_A}_{A}-(A\leftrightarrow B).~~~(I=JK,J=AB,K=CD)
\fe
Recall that the interaction currents between gauge fields and conformally coupled scalars can be written as (we still use the notation $I=JK,J=AB,K=CD$ to refer to the given deconcatenation in the interaction currents for a given binary tree):
\ie
&\mathcal{J}^{\Gamma}_{I\mu}=-k_{J\mu}\Phi^{\Gamma_J}_{J}\Phi^{\Gamma_K}_{K}+k_{K\mu}\Phi^{\Gamma_K}_{K}\Phi^{\Gamma_J}_{J}+\frac{1}{z}[\tilde{\mathcal{A}}^{\Gamma_A}_{A\mu}\Phi^{\Gamma_B}_{B}\Phi^{\Gamma_K}_{K}-\tilde{\mathcal{A}}^{\Gamma_B}_{B\mu}\Phi^{\Gamma_A}_{A}\Phi^{\Gamma_K}_{K}-\Phi^{\Gamma_J}_{J}\tilde{\mathcal{A}}^{\Gamma_C}_{C\mu}\Phi^{\Gamma_D}_{D}+\Phi^{\Gamma_J}_{J}\tilde{\mathcal{A}}^{\Gamma_D}_{D\mu}\Phi^{\Gamma_C}_{C}]\\
&\mathcal{J}^{\Gamma}_{Iz}=-i\Phi^{\Gamma_J}_{J}\overleftrightarrow{\partial}\Phi^{\Gamma_K}_{K}+\frac{1}{z}[\tilde{\alpha}^{\Gamma_A}_{A}\Phi^{\Gamma_B}_{B}\Phi^{\Gamma_K}_{K}-\tilde{\alpha}^{\Gamma_B}_{B}\Phi^{\Gamma_A}_{A}\Phi^{\Gamma_K}_{K}-\Phi^{\Gamma_J}_{J}\tilde{\alpha}^{\Gamma_C}_{C}\Phi^{\Gamma_D}_{D}+\Phi^{\Gamma_J}_{J}\tilde{\alpha}^{\Gamma_D}_{D}\Phi^{\Gamma_C}_{C}].\\
&~~~~~~~~(I=JK,J=AB,K=CD)
\fe
For $z$ component of gauge field currents $\alpha_I$, we have
\ie\label{eq:eomalphaspecificdiag}
k_{I}^{2}\alpha^{\Gamma_I}_{I}&=\frac{1}{z}[2\alpha^{\Gamma_K}_{K}(k_{K}\cdot\mathcal{A}^{\Gamma_J}_{J})+i(\mathcal{A}^{\Gamma_J}_{J}\cdot\partial_{z}\mathcal{A}^{\Gamma_K}_{K})-(J\leftrightarrow K)]\\
&~~~+\frac{1}{z^{2}}[\alpha^{\Gamma_C}_{C}(\mathcal{A}^{\Gamma_J}_{J}\cdot\mathcal{A}^{\Gamma_D}_{D})-(C\leftrightarrow D)+\alpha^{\Gamma_B}_{B}(\mathcal{A}^{\Gamma_A}_{A}\cdot\mathcal{A}^{\Gamma_K}_{K})-(A\leftrightarrow B)].\\
&~~~(I=JK,J=AB,K=CD)
\fe
Note that $\tilde{\mathcal{A}}_{I\mu},\tilde{\alpha}_{I}$ vanish for odd $I$ since such currents correspond to correlation functions with odd number scalar legs. And we also have 
\begin{equation}\label{eq:traceAA}
    \mathcal{T}^{XY}(\mathcal{A}^{\Gamma_X}_{X}\cdot\mathcal{A}^{\Gamma_Y}_{Y})=(-1)^{\frac{|X|+|Y|}{2}}(-\Phi^{\Gamma_X}_{X}\Phi^{\Gamma_Y}_{Y}+\tilde{\mathcal{A}}^{\Gamma_X}_{X}\cdot\tilde{\mathcal{A}}^{\Gamma_Y}_{Y})
\end{equation}
for arbitrary subsets $X,Y$ of $I$ with $|X|,|Y|<|I|$. This equation can be obtained from the induction ansatz directly. Substituting the ansatz \eqref{proof} for lower points and \eqref{eq:traceAA} into the equation of moition for $z$ component \eqref{eq:eomalphaspecificdiag}, we have:
\ie
\mathcal{T}^{I}k_{I}^{2}\alpha^{\Gamma}_{I}&=\frac{1}{z}[2(-1)^{\frac{|I|}{2}}\tilde{\alpha}^{\Gamma_{K}}_{K}(k_{K}\cdot\tilde{\mathcal{A}}^{\Gamma_{J}}_{J})+(-1)^{\frac{|I|}{2}}i(\tilde{\mathcal{A}}^{\Gamma_{J}}_{J}\cdot\partial_{z}\tilde{\mathcal{A}}^{\Gamma_{K}}_{K})+i(-1)^{\frac{|I|}{2}-1}\Phi_{J}^{\Gamma_{J}}\partial_{z}\Phi_{K}^{\Gamma_{K}}-(J\leftrightarrow K)]\\
&~~~+\frac{1}{z^{2}}[(-1)^{\frac{|I|}{2}}\tilde{\alpha}^{\Gamma_{C}}_{C}(\tilde{\mathcal{A}}^{\Gamma_{J}}_{J}\cdot\tilde{\mathcal{A}^{\Gamma_{D}}_{D}}-\Phi_{J}^{\Gamma_{J}}\Phi_{D}^{\Gamma_{D}})-(C\leftrightarrow D)+(-1)^{\frac{|I|}{2}}\tilde{\alpha}^{\Gamma_{B}}_{B}(\tilde{\mathcal{A}}^{\Gamma_{A}}_{A}\cdot\tilde{\mathcal{A}}^{\Gamma_{K}}_{K}-\Phi_{A}^{\Gamma_{A}}\Phi_{K}^{\Gamma_{K}})\\
&~~~-(A\leftrightarrow B)]\\
&=(-1)^{\frac{|I|}{2}}\frac{1}{z}[2\tilde{\alpha}^{\Gamma_{K}}_{K}(k_{K}\cdot\tilde{\mathcal{A}}^{\Gamma_{J}}_{J})+i(\tilde{\mathcal{A}}^{\Gamma_{J}}_{J}\cdot\partial_{z}\tilde{\mathcal{A}}^{\Gamma_{K}}_{K})-(J\leftrightarrow K)]\\
&~~~-(-1)^{\frac{|I|}{2}}\frac{1}{z^{2}}[\tilde{\alpha}^{\Gamma_{C}}_{C}(\tilde{\mathcal{A}}^{\Gamma_{J}}_{J}\cdot\tilde{\mathcal{A}^{\Gamma_{D}}_{D}})-(C\leftrightarrow D)+\tilde{\alpha}^{\Gamma_{B}}_{B}(\tilde{\mathcal{A}}^{\Gamma_{A}}_{A}\cdot\tilde{\mathcal{A}^{\Gamma_{K}}_{K}})-(A\leftrightarrow B)]+(-1)^{\frac{|I|}{2}}\frac{1}{z}\mathcal{J}^{\Gamma}_{Iz}\\
&=(-1)^{\frac{|I|}{2}}k_{I}^{2}\tilde{\alpha}_{I}^{\Gamma}.~~~(I=JK,J=AB,K=CD)
\fe
It shows that we have proved the last line of the relations \eqref{proof} we want to proof at the beginning of the subsection by induction. The other two relations can be similarly proved through the induction procedure. For the second relation in \eqref{proof}:
\ie
\mathcal{T}^{I}\frac{1}{z^{2}}(\mathcal{D}^2_{I}+d-1)\mathcal{A}^{\Gamma}_{I}\cdot v&=(-1)^{\frac{|I|}{2}}i(k_{I}\cdot v)[\partial_{z}+(2-d)/z]\tilde{\alpha}^{\Gamma}_{I}\\
&~~~+(-1)^{\frac{|I|}{2}}\frac{1}{z}[(k_{K}\cdot v\tilde{\alpha}^{\Gamma_K}_{K}+2i\partial_{z}\tilde{\mathcal{A}}^{\Gamma_K}_{K}\cdot v)\tilde{\alpha}^{\Gamma_J}_{J}+k_{K}\cdot v(\tilde{\mathcal{A}}^{\Gamma_J}_{J}\cdot\tilde{\mathcal{A}}^{\Gamma_K}_{K})\\
&~~~+\tilde{\mathcal{A}}^{\Gamma_K}_{K}\cdot v[i(\partial_{z}-d/z)\tilde{\alpha}^{\Gamma_J}_{J}-2k_{K}\cdot\tilde{\mathcal{A}}^{\Gamma_J}_{J}]-(J\leftrightarrow K)]\\
&~~~+(-1)^{\frac{|I|}{2}}\frac{1}{z^{2}}[\tilde{\alpha}^{\Gamma_J}_{J}\tilde{\alpha}^{\Gamma_C}_{C}\tilde{\mathcal{A}}^{\Gamma_D}_{D}\cdot v+(\tilde{\mathcal{A}}^{\Gamma_J}_{J}\cdot\tilde{\mathcal{A}}^{\Gamma_C}_{C})\tilde{\mathcal{A}}^{\Gamma_D}_{D}\cdot v-(C\leftrightarrow D)\\
&~~~+\tilde{\alpha}^{\Gamma_K}_{K}\tilde{\alpha}^{\Gamma_B}_{B}\tilde{\mathcal{A}}^{\Gamma_A}_{A}\cdot v+(\tilde{\mathcal{A}}^{\Gamma_B}_{B}\cdot\tilde{\mathcal{A}}^{\Gamma_K}_{K})\tilde{\mathcal{A}}^{\Gamma_A}_{A}\cdot v-(A\leftrightarrow B)]\\
&~~~-(-1)^{\frac{|I|}{2}}\frac{1}{z}v\cdot\mathcal{J}^{\Gamma}_{I}\\
&=(-1)^{\frac{|I|}{2}}\frac{1}{z^{2}}(\mathcal{D}^2_{I}+d-1)\tilde{\mathcal{A}}^{\Gamma}_{I}\cdot v.~~~(I=JK,J=AB,K=CD)
\fe
For the first line in \eqref{proof}, we have 
\ie\label{pro}
\mathcal{T}^{I}\frac{1}{z^{2}}(\mathcal{D}^2_{I}+d-1)\mathcal{A}^{\Gamma}_{I}\cdot\epsilon_{n}&=(-1)^{\frac{|I|-1}{2}}\frac{1}{z}[(2i\partial_{z}\Phi^{\Gamma_K}_{K})\tilde{\alpha}^{\Gamma_J}_{J}+\Phi^{\Gamma_K}_{K}[i(\partial_{z}-d/z)\tilde{\alpha}^{\Gamma_J}_{J}-2k_{K}\cdot\tilde{\mathcal{A}}^{\Gamma_J}_{J}]-(J\leftrightarrow K)]\\
&~~~+(-1)^{\frac{|I|-1}{2}}\frac{1}{z^{2}}[\tilde{\alpha}^{\Gamma_J}_{J}\tilde{\alpha}^{\Gamma_C}_{C}\Phi^{\Gamma_D}_{D}+(\tilde{\mathcal{A}}^{\Gamma_J}_{J}\cdot\tilde{\mathcal{A}}^{\Gamma_C}_{C})\Phi^{\Gamma_D}_{D}-(C\leftrightarrow D)+\tilde{\alpha}^{\Gamma_K}_{K}\tilde{\alpha}^{\Gamma_B}_{B}\Phi^{\Gamma_A}_{A}\\
&~~~+(\tilde{\mathcal{A}}^{\Gamma_B}_{B}\cdot\tilde{\mathcal{A}^{\Gamma_K}_{K}})\Phi^{\Gamma_A}_{A}-(A\leftrightarrow B)]\\
&=(-1)^{\frac{|I|-1}{2}}\frac{1}{z^{2}}(\mathcal{D}^2_{I}+d-1)\Phi^{\Gamma}_{I}.~~~(I=JK,J=AB,K=CD)
\fe
Note that if $J,\ C,\ D$, for example, are all odd, then it will appear the term $\Phi_{J}\Phi_{C}\Phi_{D}$ in the second line of \eqref{pro}. However, it will be canceled by the corresponding term in $(C\leftrightarrow D)$. Thus we have proven that ($\ref{proof}$) is correct for any given binary tree. Then the proof for the full BG currents and the correlation function can be straightforwad as we just need to sum over all the possible binary trees. 

In fact, by taking the flat limit, our results are consistent with the flat case in any dimension $d$, which means that we have given a semi-on-shell proof for unifying relations \eqref{cor} in flat spacetime.

\subsection{Proof of the Unifying Relations in (A)dS for the General Case}\label{more}
In flat spacetime, we can get mixed amplitudes by acting $\ma{T}[ij]$ on gluon amplitudes, like \eqref{mix}. We expect that in (A)dS, the correlation functions also have this property. This subsection will discuss the unifying relations \eqref{cor} for BG currents with mixed legs. As before, we will focus on the currents for a given binary tree in this subsection rather than the sum of all binary trees.

First, we need to show how to obtain BG currents for different types of external lines. In fact, this can be realized by modifying the initial conditions \cite{Berends:1987me}. For example, for scalar currents of minimal coupling scalar theory, we need to keep the bulk leg scalar, and the boundary legs can be given different conditions: $\tilde{\mathcal{A}}_{i}=0, \Phi_{i}\neq 0$ for scalars and $\Phi_{i}=0, \tilde{\ma{A}}_{i}\neq0$ for gluons. Then the BG currents we get will correspond to mixed correlation functions. Let us show some examples where we only consider the binary tree FIG. \ref{fig:4points}, i.e. $\Gamma=[[1,2],3]$.

The mixed scalar currents for the ``particle 2" scalar and other gluons should impose the following initial conditions:
\begin{equation}
\begin{split}
    \Phi_1&=\Phi_3=0,\\
    \tilde{\ma{A}}_2&=0.
\end{split}
\end{equation}
In other words, we do not switch off the gauge field source at the boundary and instead turn off the scalar source for ``particle 1" and ``particle 3" at the boundary. Then, the currents for these initial conditions can be calculated from the recursion relations \eqref{eq:scalarrecur}:
\begin{equation}
    \begin{split}
        (\mathcal{D}_{123}^2+d-1)\Phi^{[[1,2],3]}_{123,2}=-4z\phi_3(k_{12}\cdot\epsilon_3)\frac{1}{\ma{D}^2_{12}+d-1}z\phi_2(k_2\cdot\epsilon_1)\phi_1-(\epsilon_{1}\cdot\epsilon_{3})\phi_{1}\phi_{2}\phi_{3}.
    \end{split}
\end{equation}
Exactly the same as $\mathcal{T}[24](\mathcal{D}_{123}^2+d-1)\mathcal{A}^{[[1,2],3]}_{123}\cdot\epsilon_{4}$. 

We can also consider other initial conditions:
\begin{equation}
    \begin{split}
        \Phi_1=\Phi_2=0,\\
        \tilde{\ma{A}}_3=0.
    \end{split}
\end{equation}
Now we switch on the gauge fields source for ``particle 1" and ``particle 2" at the boundary, and ``particle 3" is a scalar. The scalar currents for such initial conditions can be written as
\begin{equation}
    \begin{split}
        (\mathcal{D}_{123}^2+d-1)\Phi^{[[1,2],3]}_{123,3}
        &=z\phi_3\frac{1}{D_{12}^2+d-1}[\frac{2z^2k_{12}\cdot k_3}{k_{12}^2}(\epsilon_1\cdot\epsilon_2)\frac{k_2^2-k_1^2}{z}\phi_1\phi_2\\
		&~~~-2z(k_2\cdot k_3-k_1\cdot k_3)(\epsilon_1\cdot\epsilon_2)\phi_1\phi_2+4z(\epsilon_2\cdot k_3)(k_2\cdot\epsilon_1)\phi_2\phi_1\\
		&~~~-4z(\epsilon_1\cdot k_3)(k_1\cdot\epsilon_2)\phi_1\phi_2]-z\phi_3\frac{1}{k_{12}^2}(\epsilon_1\cdot\epsilon_2)(\frac{k_2^2-k_1^2}{z}\phi_1\phi_2-\frac{2}{z^2}\phi_1\overleftrightarrow{\partial}\phi_2)\\
		&~~~-\frac{2}{k_{12}^2}(\epsilon_1\cdot\epsilon_2)\phi_1\overleftrightarrow{\partial}\phi_2\partial_z\phi_3.
    \end{split}
\end{equation}
This result is the same as $\ma{T}[34](\mathcal{D}_{123}^2+d-1)\ma{A}^{[[1,2],3]}_{123}\cdot\epsilon_{4}$. 

Next, we want to prove the unifying relations \eqref{pppp} for mixed currents, i.e. $X\neq I$. Here we also aim to prove that the relation \eqref{pppp} holds for any binary tree. Thus we take the deconcatenation $I=JK, J=AB, K=CD$ for a given binary tree. Let us start with a simple case $\ma{T}^{Xn}=\ma{T}[in]$ first, where $i$ is an arbitrary letter in $I$. In other words, only the $i$-th external leg is the scalar. In this case, the operator $\ma{T}^{in}$ is equal to $\ma{T}[in]$. For the BG currents whose initial conditions are $\tilde{\mathcal{A}}_{i}=0$ for a certain $i$ and all the other boundary legs are gluons, we have the following ansatz:
\ie\label{partial}
\ma{T}[in](\mathcal{D}^2_{I}+d-1)\mathcal{A}^{\Gamma}_{I}\cdot\epsilon_{n}=(\mathcal{D}^2_{I}+d-1)\Phi^{\Gamma}_{I,i},
\fe
where $\Phi_{I}^{i}$ denotes to the mixed currents with only the $i$-th particle scalar. We will prove this by induction. Still again, we focus on some arbitrary given binary tree $\Gamma$ relating to a unique deconcatenation. And we consider the following deconcatenation: $I=JK,J=AB,K=CD$. The effective terms of the left-hand side in \eqref{partial} are
\ie
(\mathcal{D}^2_{I}+d-1)\mathcal{A}^{\Gamma}_{I}\cdot\epsilon_{n}&\sim z[(2i\partial_{z}\mathcal{A}^{\Gamma_{K}}_{K}\cdot\epsilon_{n})\alpha^{\Gamma_{J}}_{J}+\mathcal{A}^{\Gamma_{K}}_{K}\cdot\epsilon_{n}[i(\partial_{z}-d/z)\alpha^{\Gamma_{J}}_{J}-2k_{K}\cdot\mathcal{A}^{\Gamma_{J}}_{J}]-(J\leftrightarrow K)]\\
&~~~+[\alpha^{\Gamma_{J}}_{J}\alpha^{\Gamma_{C}}_{C}\mathcal{A}^{\Gamma_{D}}_{D}\cdot\epsilon_{n}+(\mathcal{A}^{\Gamma_{J}}_{J}\cdot\mathcal{A}^{\Gamma_{C}}_{C})\mathcal{A}^{\Gamma_{D}}_{D}\cdot\epsilon_{n}-(C\leftrightarrow D)\\
&~~~+\alpha^{\Gamma_{K}}_{K}\alpha^{\Gamma_{B}}_{B}\mathcal{A}^{\Gamma_{A}}_{A}\cdot\epsilon_{n}+(\mathcal{A}^{\Gamma_{B}}_{B}\cdot\mathcal{A}^{\Gamma_{K}}_{K})\mathcal{A}^{\Gamma_{A}}_{A}\cdot\epsilon_{n}-(A\leftrightarrow B)].\\
&~~~(I=JK,J=AB,K=CD)
\fe
The letter $i$ must be in one of the words $A,\ B,\ C,\ D$. In other words, the non zero scalar initial condition $\Phi_i$ must be in one of the binary tree branch $A,\ B,\ C,\ D$. Different cases will let different terms survive after acting $\ma{T}[in]$. However, the analysis of all these cases are similar. Without loss of generality, we can choose the letter $i$ in the word $D$, then the effective terms are
\ie
(\mathcal{D}^2_{I}+d-1)\mathcal{A}^{\Gamma}_{I}\cdot\epsilon_{n}&\sim z[(2i\partial_{z}\mathcal{A}^{\Gamma_{K}}_{K}\cdot\epsilon_{n})\alpha^{\Gamma_{J}}_{J}+\mathcal{A}^{\Gamma_{K}}_{K}\cdot\epsilon_{n}[i(\partial_{z}-d/z)\alpha^{\Gamma_{J}}_{J}-2k_{K}\cdot\mathcal{A}^{\Gamma_{J}}_{J}]\\
&~~~+[\alpha^{\Gamma_{J}}_{J}\alpha^{\Gamma_{C}}_{C}\mathcal{A}^{\Gamma_{D}}_{D}\cdot\epsilon_{n}+(\mathcal{A}^{\Gamma_{J}}_{J}\cdot\mathcal{A}^{\Gamma_{C}}_{C})\mathcal{A}^{\Gamma_{D}}_{D}\cdot\epsilon_{n}]~~~(I=JK,J=AB,K=CD)
\fe
If for currents of scalar theory such as $\tilde{\ma{A}}_{J}$, $\tilde{\alpha}_{J}$, $\Phi_{J}$, all initial conditions are for gluons, then they are the same as the corresponding gluon currents and $\Phi_{J}$=0. Also, note that $\tilde{A}_{I\mu}^{i}$ (also $\tilde{\alpha}^{i}$) must be zero for there is no scattering process with odd number scalar external legs. Then by induction we have
\ie
\ma{T}[in](\mathcal{D}^2_{I}+d-1)\mathcal{A}^{\Gamma}_{I}\cdot\epsilon_{n}&=z[(2i\partial_{z}\Phi^{\Gamma_{K}}_{K,i})\tilde{\alpha}^{\Gamma_{J}}_{J}+\Phi_{K,i}^{\Gamma_{K}}[i(\partial_{z}-d/z)\tilde{\alpha}^{\Gamma_{J}}_{J}-2k_{K}\cdot\tilde{\mathcal{A}}^{\Gamma_{J}}_{J}]\\
&~~~+[\tilde{\alpha}^{\Gamma_{J}}_{J}\tilde{\alpha}^{\Gamma_{C}}_{C}\Phi_{D,i}^{\Gamma_{D}}+(\tilde{\mathcal{A}}^{\Gamma_{J}}_{J}\cdot\tilde{\mathcal{A}}^{\Gamma_{C}}_{C})\Phi_{D,i}^{\Gamma_{D}}]\\
&=(\mathcal{D}^2_{I}+d-1)\Phi^{\Gamma}_{I,i}.~~~(I=JK,J=AB,K=CD)
\fe

To extend these relations to the correlation functions, we also need to deal with the case of gluon currents $\tilde{A}_{I\mu}$, whose bulk leg is gluon, in the minimal coupling scalar theory. We need to prove 
\ie
\ma{T}[ij](\mathcal{D}^2_{I}+d-1)\ma{A}^{\Gamma}_{I}\cdot v=-(\mathcal{D}^2_{I}+d-1)\tilde{\ma{A}}^{\Gamma}_{I,ij}\cdot v
\fe
and
\ie
\ma{T}[ij]k_{I}^{2}\alpha^{\Gamma}_{I}=-k^{2}_{I}\tilde{\alpha}^{\Gamma}_{I,ij}.
\fe
Here the labels ``$i\ ,j$" mean that all external legs are gluons except the $i$-th and the $j$-th, and $v_{\mu}$ is an arbitrary vector. Obviously, the unifying relations hold for 3-pt currents. By induction, as the proof of \eqref{partial} we give, one can easily prove the cases for $i,\ j\in A\ (B,\ C,\ D)$ and $i\in A, j\in B\ (i\in C, j\in D)$. For the case, without loss of generality, $i\in A$ and $j\in C$, we need to use \eqref{partial} and then complete the proof.

Now we need to construct the mixed correlation functions. We define the 2-scalar-$(n-2)$-gluon correlation functions (with the first 2 particles scalar) as follows
\ie\label{mixco}
A_{\rm mixed}(1^{s},2^{s},3^{g},\cdots,n^{g})&=-\frac{1}{N}\int\frac{dz}{z^{d+1}}[-\tilde{\ma{A}}_{n}\cdot(\mathcal{D}^2_{123\cdots n-1}+d-1)\tilde{\ma{A}}_{123\cdots n-1,12}\\
&~~~+\text{cyclic}(1^{s},2^{s},3^{g},\cdots,n^{g})],
\fe
where the ``cyclic" means we sum over all the terms correspond to the legs in the bracket. Here the scalar leg $1^{s}$ corresponds to the term $\Phi_{1}(\mathcal{D}^2_{234\cdots n}+d-1)\Phi_{234\cdots n,2}$ and so as $2^{s}$, while term correspond to a certain gluon leg $i^{g}$ has the similar form as the term given in \eqref{mixco}. Of course, the currents that appear in the definition \eqref{mixco} need to sum over all possible binary trees, so there is not a upper label $\Gamma$. For example,
\ie
A_{\rm mixed}(1^{s},2^{g},3^{s},4^{g})=&-\frac{1}{N}\int\frac{dz}{z^{d+1}}[-\tilde{A}_{4}\cdot(\mathcal{D}^2_{123}+d-1)\tilde{A}_{123,13})+\Phi_{3}(\mathcal{D}^2_{124}+d-1)\Phi_{412,1}\\
&-\tilde{A}_{2}\cdot(\mathcal{D}^2_{134}+d-1)\tilde{A}_{341,13})+\Phi_{1}(\mathcal{D}^2_{234}+d-1)\Phi_{234,3}].
\fe
We need to explain the extra minus sign of the gluon current terms. The origin of this minus sign, together with the factor $(-1)^{\frac{|I|-1}{2}}$ and $(-1)^{\frac{|I|}{2}}$in subsection \ref{subsec:proof}, is that, the 2-pt gluon current of the binary tree $[1,2]$ with both boundary legs scalars has an opposite sign with the Feynman rules for the corresponding 3-pt vertex, while for the 2-pt scalar currents $[1,2]$ with one boundary leg scalar and the other gluon, the sign is the same as the Feynman rules (assuming all particles be incoming). Thus we need to multiply a extra minus signs to the former case in order to make it consistent with the Feynman rules. We also need to point out that the power of minus sign we need to multiply is only related to the number of $\ma{T}[ij]$ we act on the gluon currents where $i,j\neq n$, rather than the type of binary trees. Note that for pure scalar correlators, there will only a overall factor of minus signs; hence the definition \eqref{scacor} makes sense. However, as for our unifying relations, one must consider this extra factor.

Then we get the unifying relations for mixed amplitudes after some simple calculation with the relations for currents we have proven before:
\ie
\ma{T}[ij]A_{\rm YM\,}(1,2,3,\cdots,n)=A_{\rm mixed}(1^{g},2^{g},\cdots,i^{s},\cdots,j^{s},\cdots n^{g}).
\fe
One can prove the unifying relations as above for the case that acting two or more $\ma{T}[ij]$ on the $n$-pt currents in $d=1$ by using the methods we have used. Note that we still need to sum over the trace operators for all possible trace structures of the scalars. Now we turn to the general case. It is not hard to prove the following formulae by induction:
\ie
\ma{T}^{Xn}(\ma{D}^{2}_{I}+d-1)\ma{A}^{\Gamma}_{I}\cdot\epsilon_{n}&=(-1)^{\frac{|X|-1}{2}}(\ma{D}^{2}_{I}+d-1)\Phi^{\Gamma}_{I,X}\ \ (|X|\in \text{odd})\\
\ma{T}^{X}(\ma{D}^{2}_{I}+d-1)\ma{A}^{\Gamma}_{I}\cdot v&=(-1)^{\frac{|X|}{2}}\tilde{\ma{A}}^{\Gamma}_{I,X}\cdot v\ \ (|X|\in \text{even})\\
\ma{T}^{X}k_{I}^{2}\alpha^{\Gamma}_{I}&=(-1)^{\frac{|X|}{2}}k_{I}^{2}\tilde{\alpha}_{I,X}^{\Gamma}\ \ (|X|\in \text{even}).
\fe
As before, $v_{\mu}$ can be an arbitrary vector such as the momentum vectors and the polarization vectors, which means we have also solved the case for gluon currents and then one can construct the correlation functions. From our analysis before, there is a relative minus sign between the scalar current terms and the gluon current terms when we construct the correlators. In addition, this result is consistent with the flat case after taking the flat limit; hence one can also get the proof of the unifying relations for flat amplitudes.

It is worth saying that we expect one can use our method to prove the unifying relations for the full YMS. In YMS, scalars have two color indices, which means we can distinguish each trace structure of the scalars after double color ordering. We hope that, in this case, we do not need to sum over the trace operators at all and can obtain the unifying relations for a certain trace operator instead. We will discuss this case in our future work.

\section{Applications in Cosmology}
\label{sec:applications}
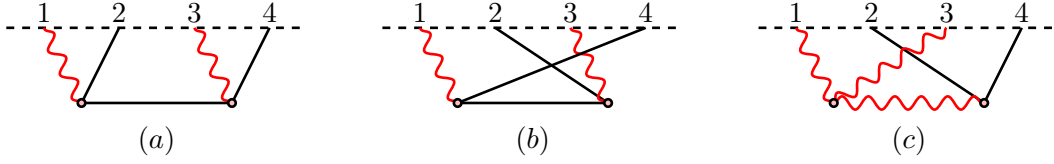
\begin{figure}
\centering
\begin{tikzpicture}[line width=1pt,scale=1]
\draw[scalarnoarrow](0,0)--(4,0);
\draw[photon](1,-1)--(0.5,0);
\draw[fermionnoarrow](1,-1)--(1.5,0);
\draw[photon](3,-1)--(2.5,0);
\draw[fermionnoarrow](3,-1)--(3.5,0);
\draw[fermionnoarrow](1,-1)--(3,-1);
\draw[fermionnoarrow,fill=pink] (1,-1) circle (.05cm);
\draw[fermionnoarrow,fill=pink] (3,-1) circle (.05cm);
\node at (2,-1.5) {$(a)$};
\node at (0.5,0.2) {1};
\node at (1.5,0.2) {2};
\node at (2.5,0.2) {3};
\node at (3.5,0.2) {4};
\begin{scope}[shift={(5,0)}]
\draw[scalarnoarrow](0,0)--(4,0);
\draw[photon](1,-1)--(0.5,0);
\draw[fermionnoarrow](3,-1)--(1.5,0);
\draw[photon](3,-1)--(2.5,0);
\draw[fermionnoarrow](1,-1)--(3.5,0);
\draw[fermionnoarrow](1,-1)--(3,-1);
\draw[fermionnoarrow,fill=pink] (1,-1) circle (.05cm);
\draw[fermionnoarrow,fill=pink] (3,-1) circle (.05cm);
\node at (2,-1.5) {$(b)$};
\node at (0.5,0.2) {1};
\node at (1.5,0.2) {2};
\node at (2.5,0.2) {3};
\node at (3.5,0.2) {4};
\end{scope}
\begin{scope}[shift={(10,0)}]
\draw[scalarnoarrow](0,0)--(4,0);
\draw[photon](1,-1)--(0.5,0);
\draw[fermionnoarrow](3,-1)--(1.5,0);
\draw[photon](1,-1)--(2.5,0);
\draw[fermionnoarrow](3,-1)--(3.5,0);
\draw[photon](1,-1)--(3,-1);
\draw[fermionnoarrow,fill=pink] (1,-1) circle (.05cm);
\draw[fermionnoarrow,fill=pink] (3,-1) circle (.05cm);
\node at (2,-1.5) {$(c)$};
\node at (0.5,0.2) {1};
\node at (1.5,0.2) {2};
\node at (2.5,0.2) {3};
\node at (3.5,0.2) {4};
\end{scope}
\end{tikzpicture}
\caption{Witten diagrams for 4-pt correlation function between inflatons and gauge fields. The dashed line is the future boundary of dS spacetime, and the solid line represents the inflaton propagators, while the wavy line is the gauge field propagators. Each pink dot represents the conformally coupled vertex. $(a)$, $(b)$ and $(c)$ represent for the $s$, $t$ and $u$-channel contribution.}
\label{fig:JJOO}
\end{figure}

This section considers the applications for unifying relations in dS spacetime. During the period of inflation, our universe is dominated by dark energy, which means the primordial universe is dS spacetime. The interaction between particles and scalar fields has driven the cosmic expansion in the early universe will result in the anisotropy of Cosmic Microwave Background (CMB) at present. Therefore, the correlation functions between inflatons and other fields become a key to figuring out the history of our primordial universe. In this section, we use the propagators of dS spacetime and change AdS to dS by the method we have mentioned in subsection \ref{BGAdS}. For simplicity, we will take $l=1$ and choose $d=3$ but the generalization to arbitrary spacetime dimension is straightforward. Now we give some concrete examples to see the applications of the unifying relations in cosmology. In this subsection, we will focus on the 4-pt correlators $\la JOJO\ra$ (see FIG. \ref{fig:JJOO}).

Before the calculation, we can roughly mention the whole procedure using the recursive BG currents and the unifying relations to calculate the cosmic correlators. Fisrt, we should write down the equation of motion for gauge field including the interaction terms then calculate the relative BG currents through the perturbiner expansion ansatz. Then, we can use the BG currents to construct the pure gluons correlation functions. Finally, we write down the trace operator $\ma{T}[ij]$ where $i,j$ are those propagators we would like to set to be scalars. For example, $\ma{T}[12]$ means that we want such legs with momentum $k_1$ and $k_2$ to be scalars. Then, we obtain the cosmological correlation functions we are interested in.

In the following discussion, we consider the 4-pt correlators $\la JOJO\ra$ calculation. According to the calculation procedure we outlined above. We first calculate the 4-pt gluon correlation function through the BG currents. Plugging in the 3-pt gluon BG current \eqref{eq:3ptgluoncurrent1} into tree-level correlation function \eqref{eq:ymcorrelation}, we can write down the 4-pt gluon correlation function. Remind that the 4-pt correlation function is constructed from 3-pt BG current in AdS case.
\begin{equation}
    A_{\rm YM\,}(1,2,3,4)=-\frac{1}{4}\int\frac{dz}{z^{4}}\mathcal{A}_4\cdot(\ma{D}_{123}^2+2)\ma{A}_{123}+\text{cyclic}(1,2,3,4).
\end{equation}
Here $\ma{A}_{123}$ is the 3-pt gluon BG current which we have presented in \eqref{eq:3ptgluoncurrent1} and $\ma{A}_4$ is the single point current for gluons. We should remind again that each Feynman diagram has a unique corresponding with trace operator $\ma{T}[ij]$ because $\ma{T}[ij]$ will not change the types of channels of the correlators. For example, the $s$-channel of 4-pt correlation function $\la JOJO\ra$  is related to the $s$-channel of $\la JJJJ\ra$ by acting $\ma{T}[24]$. In other words, we can get the mixed correlation function $s$-channel through the trace operator after changing the radial coordinate to the dS one:
\ie
\ma{T}[24]A_{\rm YM\,}(1234)_{s}&=-2i(k_{2}\cdot\epsilon_{1})(k_{4}\cdot\epsilon_{3})\int\frac{d\eta}{\eta^{4}}(\eta\phi_{1}\phi_{2}\frac{1}{\ma{D}^{2}_{34}+2}\eta\phi_{3}\phi_{4}+\eta\phi_{3}\phi_{4}\frac{1}{\ma{D}^{2}_{12}+2}\eta\phi_{1}\phi_{2})\\
&=\frac{-4(k_{2}\cdot\epsilon_{1})(k_{4}\cdot\epsilon_{3})}{(|k_{1}|+|k_{2}|+s)(|k_{3}|+|k_{4}|+s)(|k_{1}|+|k_{2}|+|k_{3}|+|k_{4}|)}\\
&=\la JOJO\ra_{s}.
\fe
Here we have stripped the contact terms, and $s$ represents the $s$-channel ($s^{2}=(k_1+k_2)^{2}$). For $u$-channel (see FIG. \ref{fig:JJOO} $(c)$), there are more subtleties. In fact, this diagram is equivalent to $\ma{T}[24]A_{\rm YM\,}(1324)$, so we can calculate $\ma{T}[34]A_{\rm YM\,}(1234)$ first and then let 2 and 3 exchange. First, we have
\ie\label{eq:JJOOs}
\ma{T}[34]A_{\rm YM\,}(1234)_{s}&=-\frac{i}{k_{12}^{2}}(\epsilon_{1}\cdot\epsilon_{2})\int\frac{d\eta}{\eta^{4}}\phi_{1}\overleftrightarrow{\partial}\phi_{2}\phi_{3}\overleftrightarrow{\partial}\phi_{4}-\frac{i}{2}[(\epsilon_{1}\cdot\epsilon_{2})\frac{(k_{1}^{2}-k_{2}^{2})(k_{3}^{2}-k_{4}^{2})}{k_{12}^{2}}\\
&~~~+(\epsilon_{1}\cdot\epsilon_{2})(k_{3}-k_{4})\cdot(k_{1}-k_{2})+4(k_{3}\cdot\epsilon_{2})(k_{2}\cdot\epsilon_{1})\\
&~~~-4(k_{3}\cdot\epsilon_{1})(k_{1}\cdot\epsilon_{2})]\int\frac{d\eta}{\eta^{4}}(\eta\phi_{1}\phi_{2}\frac{1}{\ma{D}^{2}_{34}+2}\eta\phi_{3}\phi_{4}+\eta\phi_{3}\phi_{4}\frac{1}{\ma{D}^{2}_{12}+2}\eta\phi_{1}\phi_{2})\\
&=\frac{(\epsilon_{1}\cdot\epsilon_{2})(|k_{1}|-|k_{2}|)(|k_{3}|-|k_{4}|)}{k_{12}^{2}(|k_{1}|+|k_{2}|+|k_{3}|+|k_{4}|)}-[(\epsilon_{1}\cdot\epsilon_{2})\frac{(k_{1}^{2}-k_{2}^{2})(k_{3}^{2}-k_{4}^{2})}{k_{12}^{2}}\\
&~~~+(\epsilon_{1}\cdot\epsilon_{2})(k_{3}-k_{4})\cdot(k_{1}-k_{2})+4(k_{3}\cdot\epsilon_{2})(k_{2}\cdot\epsilon_{1})-4(k_{3}\cdot\epsilon_{1})(k_{1}\cdot\epsilon_{2})]\\
&~~~\times\frac{1}{(|k_{1}|+|k_{2}|+s)(|k_{3}|+|k_{4}|+s)(|k_{1}|+|k_{2}|+|k_{3}|+|k_{4}|)}.
\fe
Then we exchange 2 and 3 to get the $u$-channel mixed correlation function
\ie
\la JOJO\ra_{u}=&\frac{(\epsilon_{1}\cdot\epsilon_{3})(|k_{1}|-|k_{3}|)(|k_{2}|-|k_{4}|)}{k_{13}^{2}(|k_{1}|+|k_{2}|+|k_{3}|+|k_{4}|)}-[(\epsilon_{1}\cdot\epsilon_{3})\frac{(k_{1}^{2}-k_{3}^{2})(k_{2}^{2}-k_{4}^{2})}{k_{13}^{2}}\\
&+(\epsilon_{1}\cdot\epsilon_{3})(k_{2}-k_{4})\cdot(k_{1}-k_{3})+4(k_{2}\cdot\epsilon_{3})(k_{3}\cdot\epsilon_{1})-4(k_{2}\cdot\epsilon_{1})(k_{1}\cdot\epsilon_{3})]\\
&\times\frac{1}{(|k_{1}|+|k_{3}|+u)(|k_{2}|+|k_{4}|+u)(|k_{1}|+|k_{2}|+|k_{3}|+|k_{4}|)}.
\fe
where $u^2=(k_{1}+k_{3})^2$. We should emphasize that our results match with the results from cosmological bootstrap \cite{Baumann:2020dch}. 

It is valuable to comment more about the applications for unifying relations in cosmology. As we have presented above, in 4-pt correlation functions, unifying relations can exactly reproduce the results from cosmological bootstrap \cite{Baumann:2020dch}. Therefore, we can expect that the unifying relations can help us calculate the higher points correlation functions in the dS background. Besides, our calculations can also be applied to higher dimension background spacetime.

\section{Conclusion and Outlook}
The remarkable unifying relations in flat spacetime show a connection between amplitudes in different theories. However, unlike the proof of unifying relations in the flat case where we use factorization, here in (A)dS, we use BG currents to study the unifying relations. 
We expect the unifying relations \eqref{uni} and \eqref{uni2} can still hold in (A)dS. However, as an intuitive discussion, we only focus on a special case \eqref{cor} as a corollary of the unifying relations between the YMS and YM, which is also the more concerned case in cosmology. The discussion of the original unifying relations \eqref{uni} and \eqref{uni2} will be put in future work.

The proof of the unifying relation \eqref{cor} in (A)dS case can be reached by proving the relation \eqref{pppp}. We first consider the case that all boundary legs of the currents are the same, and explicitly calculate the 3-pt (\ref{subsec:4points}) and 5-pt (\ref{subsec:6points}) currents for given binary trees in both gauge and scalar theories. Then we prove that after the action of some trace operators $\ma{T}[ij]$, the currents of two theories are exactly the same. Similarly, we also consider the case that currents have different types of boundary legs and complete the proof of \eqref{pppp}. It is not hard to find that \eqref{cor} also holds in (A)dS case from \eqref{pppp} by summing over the currents for all possible binary trees.


We explore some applications in cosmology and we used the unifying relations \eqref{cor} to calculate the 4-pt cosmological correlators. The calculation steps can be organized as follows. First, we should use the perturbiner method to derive the gauge fields BG currents from equation of motion. Second, we can use the BG currents to construct the pure gauge correlation functions. Finally, we can write down the trace operators in unifying relations and act the trace operators on the pure gauge correlation functions to get the mixed correlation functions. We note that the recursive BG currents can help us to calculate the higher points correlation function conveniently. This may provide potential applications in future higher points correlation function calculations.

In summary, we have discussed a type of unifying relations for BG currents in (A)dS. Moreover, we show some potential applications for our unifying relations in cosmological correlators calculation. However, it is desirable to mention that there are still ambiguities we have not illustrated well. We discuss only one type of unifying relations in this work. In principle, there are still several unifying relations for which we do not give rigorous proof, which may be discussed in future work. It is necessary to mention that the correlation functions at the loop level can also be constructed by BG currents \cite{Gomez:2022dzk}. And we are looking forward to deriving the relations among 1-loop level correlation functions \cite{Zhou:2021kzv,Zhou:2022djx} in future work.

\label{sec:conclusion}

\section{Acknowledgements}
We would like to thank Chi-Ming Chang and Yi-Jian Du for valuable discussions and useful comments on the draft. YT is partly supported by National Key R\&D Program of China (NO. 2020YFA0713000). QC is partly supported by National Key R\&D Program of China (Grant No. 2017YFA0402204).

\section{Appendix}
This appendix focuses on the analytical calculation of the time integral in the mixed correlation function. Before the explicit calculation, let us explain the notations in the correlation functions. And we should also note that in the following discussion since we only focus on the time integral where we do not talk about the recursions, the notations for momentum do not cause any ambiguities. Thus we denote $|k|$ as $k$ for simplicity. First, the inverse of the d'Alembert operator is defined as 
\begin{equation}
(\ma{D}_I^2-M^2)^{-1}\ma{O}(\eta)=\int\frac{d\eta'}{\eta'^4}G_I(\eta,\eta')\ma{O}(\eta'),
\end{equation}
where $\ma{O}$ is a arbitrary operator and $G_I(\eta,\eta')$ is the bulk-to-bulk propagator which satisfy the following equation:
\begin{equation}\label{eq:propagator}
    (\ma{D}_I^2-M^2)G_I=\eta^{4}\delta(\eta-\eta').
\end{equation}
In section \ref{sec:applications}, we need to evaluate the bulk-to-bulk propagator for gluon at $d=3$. When we are ready to solve the bulk-to-bulk propagator, we should impose some initial conditions. In our following calculation, we impose the Bunch-Davis (BD) vacuum, representing a non-particle state at the past infinity. Thus, the bulk-to-bulk propagator for color-stripped gluon can be written as \cite{Gomez:2021ujt}
\begin{equation}
\begin{split}
    G_{k_s}(\eta_1,\eta_2)&=-\frac{\pi  (\eta_1 \eta_2)^{3/2} }{4i }\left(H_{\frac{1}{2}}^{(2)}(-k_s \eta_1) H_{\frac{1}{2}}^{(2)}(-k_s \eta_2)+H_{\frac{1}{2}}^{(1)}(-k_s \eta_1) H_{\frac{1}{2}}^{(2)}(-k_s \eta_2)\right)\theta(\eta_1-\eta_2)\\
    &-\frac{\pi  (\eta_1 \eta_2)^{3/2} }{4i }\left(H_{\frac{1}{2}}^{(2)}(-k_s \eta_2) H_{\frac{1}{2}}^{(2)}(-k_s \eta_1)+H_{\frac{1}{2}}^{(1)}(-k_s \eta_2) H_{\frac{1}{2}}^{(2)}(-k_s \eta_1)\right)\theta(\eta_2-\eta_1),
\end{split}
\end{equation}
where $s$ denotes the $s$ channel and $H^{(1)}$, $H^{(2)}$ are separately the first and second type Hankel function. Recall that the equation of motion for conformally coupled scalars is
\begin{equation}
    (\ma{D}_k^2+2)\phi(\eta,k)=0.
\end{equation}
Also, impose the BD vacuum, we can write down the mode function for conformally coupled scalars:
\begin{equation}
    \phi(\eta,k)=-\sqrt{\frac{\pi}{2}}(-\eta)^{3/2}k^{1/2}H_{\frac{1}{2}}^{(2)}(-k \eta).
\end{equation}
Now we can carry out the time integral in the mixed correlation function. First, in $s$-channel, we have to evaluate the following integral:
\begin{equation}
    \begin{split}
        \ma{I}_s&=\int\frac{d\eta_1}{\eta_1^3}\phi(\eta_1,k_1)\phi(\eta_1,k_2)\frac{1}{\ma{D}_{34}^2+2}\eta_2\phi(\eta_2,k_3)\phi(\eta_2,k_4)\\
        &=\int\frac{d\eta_1}{\eta_1^3}\frac{d\eta_2}{\eta_2^3}\phi(\eta_1,k_1)\phi(\eta_1,k_2)G_{k_s}(\eta_1,\eta_2)\phi(\eta_2,k_3)\phi(\eta_2,k_4).
    \end{split}
\end{equation}
Note that in the propagator, we have time ordering dependence. We consider the two different time-ordering separately, and for simplicity but without loss of any generality, we focus on the positive time ordering, $\eta_1>\eta_2$. Then the time integral can be carried out analytically:
\begin{equation}
    \begin{split}
        \ma{I}_{s+}&=-\int\frac{d\eta_1}{\eta_1^3}\frac{d\eta_2}{\eta_2^3}\frac{\pi^2}{4}(\eta_1\eta_2)^3\sqrt{k_1k_2k_3k_4}H_{\frac{1}{2}}^{(2)}(-k_1 \eta_1)H_{\frac{1}{2}}^{(2)}(-k_2 \eta_1)H_{\frac{1}{2}}^{(2)}(-k_3 \eta_2)H_{\frac{1}{2}}^{(2)}(-k_4 \eta_2)\\
        &~~~\times \frac{\pi  (\eta_1 \eta_2)^{3/2} }{4 i}\left(H_{\frac{1}{2}}^{(2)}(-k_s \eta_1) H_{\frac{1}{2}}^{(2)}(-k_s \eta_2)+H_{\frac{1}{2}}^{(1)}(-k_s \eta_1) H_{\frac{1}{2}}^{(2)}(-k_s \eta_2)\right)\theta(\eta_1-\eta_2)\\
        &=\frac{-i}{(k_1+k_2+k_3+k_4)(k_3+k_4+k_s)(k_1+k_2+k_3+k_4+2k_s)}.
    \end{split}
\end{equation}
Together with the negative time ordering contribution, we can obtain the full integral result:
\begin{equation}
    \ma{I}_s=\frac{-i}{(k_1+k_2+k_3+k_4)(k_1+k_2+k_s)(k_3+k_4+k_s)}.
\end{equation}

The second time integral is much simpler, which looks like the contact interaction and does not have the bulk-to-bulk propagator contribution.
\begin{equation}
\begin{split}
    \ma{I}_c&=\int\frac{d\eta}{\eta^{4}}\phi_{1}\overleftrightarrow{\partial}\phi_{2}\phi_{3}\overleftrightarrow{\partial}\phi_{4}\\
    &=\int\frac{d\eta}{\eta^4}[\phi(\eta,k_1)\partial_{\eta}\phi(\eta,k_2)-\phi(\eta,k_2)\partial_{\eta}\phi(\eta,k_1)][\phi(\eta,k_3)\partial_{\eta}\phi(\eta,k_4)-\phi(\eta,k_4)\partial_{\eta}\phi(\eta,k_3)]\\
    &=\frac{i(k_1-k_2)(k_3-k_4)}{k_1+k_2+k_3+k_4}.
\end{split}
\end{equation}

\bibliographystyle{JHEP}
\bibliography{Unifying}

\end{document}